\documentclass[12pt]{iopart}

\usepackage{graphicx}       % Include figure files
\usepackage{dcolumn}        % Align table columns on decimal point
\usepackage{bm}             % bold math
\usepackage{amstext}
\usepackage{amssymb}
\usepackage{amsbsy}
\usepackage{amscd}
\usepackage{amsfonts}

\newcommand{\nn}{\nonumber}

\newcommand{\omco}{\varpi}
\newcommand{\omnO}{\omco^{(0)}}
\newcommand{\omcon}{\omco^{(n)}}
\newcommand{\SnO}{S^{(0)}}
\newcommand{\Sn}{S^{(n)}}

\newcommand{\lamcon}{\lambda^{(n)}}

\newcommand{\Tn}{T^{(n)}}

\newcommand{\rstar}{r_\ast}
\newcommand{\lam}{L}

\newcommand{\Bef}{\mathcal{B}_{ln}}

\newcommand{\Aout}{A^\text{(out)}_{l\omega}}
\newcommand{\Aoutln}{A^{(\text{out})}_{ln}}
\newcommand{\Ain}{A^\text{(in)}_{l\omega}}

\newcommand{\Cout}{C_{\text{out}}}
\newcommand{\Cin}{C_\text{in}}

\newcommand{\beq}{\begin{equation}}
\newcommand{\eeq}{\end{equation}}

\newcommand{\uin}{u_{l\omega}^{\text{in}}}

\begin{document}

%\preprint{}

\title{On an Expansion Method for Black Hole Quasinormal Modes and Regge Poles}

\author{Sam R. Dolan}
\ead{sam.dolan@ucd.ie}
\address{Complex and Adaptive Systems Laboratory / School of Mathematical Sciences, University College Dublin, Belfield, Dublin 4, Ireland}

\author{Adrian C. Ottewill}
\ead{adrian.ottewill@ucd.ie}
\address{Complex and Adaptive Systems Laboratory / School of Mathematical Sciences,  University College Dublin, Belfield, Dublin 4, Ireland}

\date{\today}
 
\begin{abstract}
We present a new method for determining the frequencies and wavefunctions of black hole quasinormal modes (QNMs) and Regge poles. The key idea is a novel ansatz for the wavefunction, which relates the high-$l$ wavefunctions to null geodesics which start at infinity and end in perpetual orbit on the photon sphere. Our ansatz leads naturally to the expansion of QNMs in inverse powers of $L = l + 1/2$ (in 4D), and to the expansion of Regge poles in inverse powers of $\omega$. The expansions can be taken to high orders. We begin by applying the method to the Schwarzschild spacetime, and validate our results against existing numerical and WKB methods. Next, we generalise the method to treat static spherically-symmetric spacetimes of arbitrary spatial dimension. We confirm that, at lowest order, the real and imaginary components of the QNM frequency are related to the orbital frequency and the Lyapunov exponent for geodesics at the unstable orbit. We apply the method to five spacetimes of current interest, and conclude with a discussion of the advantages and limitations of the new approach, and its practical applications.
\end{abstract}

\pacs{04.70.Bw, 04.30.Nk}
\maketitle

\section{Introduction}

Quasinormal modes (QNMs) are damped resonances which play a key role in black hole dynamics \cite{Kokkotas-Schmidt, Nollert, Ferrari-Gualtieri, Berti-Cardoso-Starinets}. It was noted four decades ago \cite{Vishveshwara} that a perturbed black hole will, after an initial response, emit radiation with well-defined frequencies and rates of damping, in the manner of a bell sounding its last dying pure notes \cite{Chandrasekhar}. Physically, a black hole quasinormal mode is a decaying resonance which satisfies a pair of causally-motivated boundary conditions (typically being ingoing at the event horizon, and outgoing at spatial infinity). Mathematically, `quasinormal ringing' emerges from a sum of residues of poles of the Green function in the complex frequency domain \cite{Leaver-1986}. Each pole corresponds to a quasinormal mode of a single complex frequency.  The real part of the frequency corresponds to the oscillation rate and the (negative) imaginary part corresponds to the damping rate.

In a spherically symmetric spacetime (such as the Schwarzschild spacetime), the QNM frequencies $\omega_{ln}$ are labelled by two integers: multipole $l$ and overtone $n \ge 0$. The properties of the QNM spectrum depends only on the properties of the field (e.g. spin $s$) and the black hole geometry (e.g. mass $M$, charge $Q$, angular momentum $J$).

Even after years of study, black hole QNMs retain a certain mystique. The frequency spectrum has a beautiful and ornate structure (see e.g.~Fig.~5 in \cite{Nollert}). The asymptotic properties of the highly-damped frequencies of the Schwarzschild hole have inspired speculation that QNMs may be linked to, e.g., Hawking radiation \cite{York}, black hole entropy \cite{Hod}, and loop quantum gravity \cite{Dreyer} (however, subsequent study of charged, rotating and higher-dimensional black holes has shown that any such links cannot be as simple as originally envisaged \cite{Natario-Schiappa, Domagala-Lewandowski}). QNMs are also of relevance to holographic principles such as the AdS/CFT conjecture \cite{Nunez-Starinets, Evans-Threlfall}. Here we focus merely on the role that QNMs play in wave propagation on black hole spacetimes, in the weakly-damped regime $l \gtrsim 2$, $l \gtrsim n$.

It has been known for many years that quasinormal modes are intimately linked to the existence and properties of unstable null orbits \cite{Goebel, Ferrari-Mashhoon, Mashhoon}. In the eikonal regime ($l \gg n$), the real part of the complex QNM frequency is related to the angular frequency of the orbit $\Omega$. It was recently demonstrated \cite{Lyapunov1} that, in any static spherically symmetric asymptotically flat spacetime, to leading order in $l$ the QNM frequency is
\beq
\lim_{l \rightarrow \infty} \text{Re} ( \omega_{ln} ) =  \Omega l , \quad \quad \quad \quad \lim_{l \rightarrow \infty} \text{Im} ( \omega_{ln} )  = - (n + 1/2) |\lambda|  , \label{leading-order}
\eeq
where $\lambda$ is the so-called \emph{Lyapunov exponent}: the (inverse of the) instability timescale of the unstable orbit \cite{Lyapunov2, Lyapunov3}. Motivated by this idea, we show here that the QNM frequency for spherically-symmetric black holes may be written as an expansion in inverse powers of $L = l+1/2$, as
$
\omega_{ln}  =   \omcon_{-1} L + \omcon_0 + \omcon_1 L^{-1} + \omcon_2 L^{-2} + \ldots   %\label{om-expansion}
$
We develop a simple method for determining the expansion coefficients $\{\omcon_k\}$ which may be taken to very high order. A key advantage of the method is that it provides expansions (in powers of $L^{-1}$) for the radial wavefunctions, which are well-defined over all radii $r > r_h$. It is straightforward to adapt the method to find an expansion of the Regge poles \cite{Decanini-Folacci-Jensen, Decanini-Folacci-2009} in powers of $\omega$, as we will show.

%By combining the method with standard WKB methods, we obtain leading-order (in $L$) approximations for the so-called `QNM Excitation Factors' (see \cite{Leaver-1986, Andersson-1997, Berti-Cardoso-2006}). This allows a study of the QNM decomposition of the Green function on black hole spacetimes. In particular, we may identify the singular part of the Green function (arising from the large-$L$ asymptotics of QNMs) associated with the light cone. We examine the form of the Green function close to the self-intersections of the light cone called caustics.

This paper is structured as follows. In Sec.~\ref{sec-existing-methods} we briefly review the existing methods for computing QNM spectra. In Sec.~\ref{sec-schwarzschild} we introduce the expansion-in-$L$ method by  application to the Schwarzschild spacetime, and validate against existing numerical and asymptotic results. In Sec.~\ref{sec-regge} we adapt the method to find Regge poles of the Schwarzschild spacetime. In Sec.~\ref{sec-ss} we generalize the QNM analysis to static spherically-symmetric spacetimes of arbitrary spatial dimension, and confirm the result (\ref{leading-order}). In Sec.~\ref{sec-examples} we apply our method to five spacetimes of physical interest (Reissner-Nordstr\"om; Schwarzschild-de Sitter; Nariai; Schwarzschild-Tangherlini; the canonical acoustic hole). We conclude in Sec.~\ref{sec-discussion} with a discussion of interesting applications of the method %; possible generalisation to the Kerr spacetime; 
and work for the future.
%In Sec.~\ref{} we discuss the decomposition of the scalar Green function into QNMs, and the contribution of QNMs to the singular part of the Green function on the light cone. 

%The null orbits on black hole spacetimes studied here are unstable, hence they spiral outwards to infinity, or inwards to the horizon. 

\section{Existing Methods for Determining QNMs\label{sec-existing-methods}}
A wide range of numerical methods have been developed for determining QNMs. The following non-exhaustive classification was outlined in \cite{Berti}: 
(i)~\emph{Time domain methods}, whereby the least-damped frequencies are extracted from long, stable time evolutions \cite{Vishveshwara, Dorband};
(ii)~\emph{Direct integration in the frequency domain} as pioneered in \cite{Chandrasekhar-Detweiler}; 
(iii)~\emph{Inverse potential methods} whereby the `exact' potential is replaced with a simple potential whose spectrum is known \cite{Blome-Mashhoon}; 
(iv)~\emph{WKB methods} adapted from standard techniques in quantum mechanics \cite{WKB, Iyer-Will, Iyer-1987, Konoplya};
(v)~\emph{Phase-integral methods} using integration along anti-Stokes lines in the complex-$r$ plane \cite{Phase-integral1, Phase-integral2, Natario-Schiappa};
(vi)~\emph{Continued fraction methods}, first employed in the 1930s to find electronic spectrum of the H${}^+$ ion, and adapted to black holes by Leaver \cite{Leaver-1985}.   
The relative advantages and disadvantages of these methods are discussed in several review articles \cite{Kokkotas-Schmidt, Nollert, Berti-Cardoso-Starinets}.

For numerical calculations, the continued fraction method (vi) is fast, reliable and accurate, and with certain modifications \cite{Nollert-1993}, is robust at large overtones $n$ and angular momenta $l$. However, the method itself provides relatively little physical insight. In contrast, WKB methods \cite{WKB, Iyer-Will, Iyer-1987, Konoplya} provide insight at the expense of accuracy and applicability. For example, the Bohr-Sommerfeld rule `explains' the link between the low-overtone frequencies and the shape of the peak of the radial potential \cite{WKB, Iyer-Will}. For improved accuracy, the WKB method may be extended to sixth order \cite{Konoplya} and above; nevertheless it still breaks down at large overtone $n \gg l$ or if the potential is complex. 

In this work we develop a new method which bears some passing resemblance to the WKB approach \cite{WKB, Iyer-Will, Iyer-1987, Konoplya}. However, instead of matching together independent solutions across a (real) potential barrier \cite{BenderOrzag}, we instead seek a single solution that is valid everywhere outside the horizon. It turns out that, once the correct geometrically-motivated ansatz is employed, the wavefunctions and frequencies may be expressed as expansions in the angular momentum parameter
\beq
L = l + 1/2 .
\eeq
%The ansatz is motivated by a simple line of geometrical reasoning, given later.

\section{The Expansion of Schwarzschild QNMs\label{sec-schwarzschild}}
Without further ado, let us introduce the method by applying it to the Schwarzschild spacetime, perturbations of which are governed by the `master' equation
\beq
\left[ \frac{d^2}{d \rstar^2} + \omega^2 - f(r) \left( \frac{(\lam^2-1/4)}{r^2} + \frac{2\beta}{r^3} \right) \right] u_{l\omega}(r) = 0,   \label{rad-eq-1}
\eeq
where $f(r) = 1-2/r$ and $\rstar = r + 2  \ln ( r/2 - 1)$ and $\beta = 1-s^2$ with $|s| = 0, 1, 2$ for scalar, electromagnetic and gravitational perturbations \cite{Chandrasekhar}, respectively. Note that we have fixed the black hole mass to $M=1$. Solutions which are `ingoing' at the horizon (as $\rstar \rightarrow -\infty$) satisfy
\beq
\uin(r) \sim 
\left\{ \begin{array}{ll} e^{-i \omega \rstar}, &  \rstar \rightarrow -\infty ,  \\ 
  \Ain e^{-i \omega \rstar} + \Aout e^{i \omega \rstar},   &    \rstar \rightarrow +\infty . \end{array} \right.
  \label{uin-def}
\eeq
Quasinormal modes correspond to the (complex) frequencies $\omega_{ln}$ at which $\Ain = 0$.

\subsection{Method}
Let us introduce the following ansatz
\beq
u_{l \omega}(r) = \exp \left( i \omega \int^{\rstar} \left( 1+ \frac{6}{r'} \right)^{1/2} \left( 1-\frac{3}{r'} \right) \, d\rstar' \right) v_{l \omega}(r)   \label{ansatz}
\eeq
Note that the integrand in the exponent goes as $1 + \mathcal{O}(r^{-2})$ as $\rstar \rightarrow \infty$, and as $-1 + \mathcal{O}(r-2)$ as $\rstar \rightarrow -\infty$. Hence QNM boundary conditions are automatically satisfied if $v_{l\omega}(r)$ is well-behaved in both limits. The geometric motivation behind this ansatz is explored in Sec.~\ref{sec-ss}.

Substitution of ansatz (\ref{ansatz}) into (\ref{rad-eq-1}) leads immediately to
\begin{eqnarray}
\frac{d}{dr} \left( f(r) \frac{d v}{d r} \right) + \left[ 2i\omega \left( 1+ \frac{6}{r} \right)^{1/2} \left(1 - \frac{3}{r} \right) \right] \frac{d v}{d r}   \nn
+ \\
 \left[ \frac{27 \omega^2 - \lam^2}{r^2} + \frac{27i \omega}{r^3} \left(1 + \frac{6}{r} \right)^{-1/2} + \frac{1}{4r^2} - \frac{2\beta}{r^3} \right] v &=& 0   \label{rad-eq-2}
\end{eqnarray}
At first glance, (\ref{rad-eq-2}) looks more complicated than (\ref{rad-eq-1}), but it has the advantage of being amenable to an expansion in inverse powers of $\lam$. To look for the `fundamental' (least-damped) $n=0$ mode, let us try the expansion
\begin{eqnarray}
\omega_{l,n=0} &=& L \omnO_{-1} + \omnO_0 + L^{-1} \omnO_1 + L^{-2} \omnO_2 \ldots   \label{omega-expansion}  \\
v_{l\omega}(r) &=& \exp\left( \SnO_0(r) + L^{-1} \SnO_1(r) + L^{-2} \SnO_2(r)  + L^{-3} \SnO_3(r) + \ldots \right)   \label{v-expansion}
\end{eqnarray}
Upon substituting into (\ref{rad-eq-2}), and collecting together like powers of $L$, we obtain a set of independent equations which may be used to determine the coefficients $\omco_k$ and the radial functions $S_k(r)$ (briefly dropping the superscript for brevity). We find
\begin{eqnarray}
\fl \lam^{2} : \quad \quad & & 27 \omco_{-1}^2 - 1 = 0 \quad \quad \quad \quad \Rightarrow \omco_{-1} = \pm 1 / \sqrt{27} \\
\fl \lam^{1} : \quad \quad &  & 2i \omco_{-1} \left(1 + \frac{6}{r} \right)^{1/2}\left(1 - \frac{3}{r} \right)S_0'  + \frac{54 \omco_{-1} \omco_0}{r^2} + \frac{27 i \omco_{-1}}{r^3}\left(1 + \frac{6}{r} \right)^{-1/2}  = 0  \label{lam1-eq} \\
\fl \lam^{0} : \quad \quad & & 2i  \left(1 + \frac{6}{r} \right)^{1/2}\left(1 - \frac{3}{r} \right)(\omco_{-1} S_1' + \omco_0 S_0' )  + f(S_0'' + (S_0')^2) + f' S_0' + \frac{27(2\omco_{-1}\omco_1 + \omco_0^2)}{r^2} \nn \\ 
& & + \frac{27i \omco_0}{r^3}\left(1+\frac{6}{r}\right)^{-1/2} + \frac{1}{4r^2} - \frac{2 \beta}{r^3} = 0 \\
\fl L^{-1}: \quad \quad & & \ldots
\end{eqnarray}
etc., where ${}^\prime$ denotes differentiation with respect to $r$. The QNM frequencies are found by imposing a \emph{continuity condition} on $\left\{ S^\prime_k(r) \right\}$ at the unstable null orbit radius $r = 3$. That is, we demand that $S^\prime_0$ is continuous and differentiable at $r=3$ which implies that %(\ref{lam1-eq})
\beq
\omnO_0 = \frac{-i}{2\sqrt{27}} ,
\eeq
and 
\beq
\frac{d \SnO_0}{d r} = \frac{\sqrt{27}}{2(r+6)(r-3)} \left[ \left(1+\frac{6}{r}\right)^{1/2} - \frac{\sqrt{27}}{r} \right] . 
\eeq

 % It also suggests we have the correct phase through ansatz (\ref{ansatz}), up to order $\mathcal{O}(\lam^{-1})$.

We may continue in this way, iteratively. First, we write down the equation for the $(1-k)$th power of $\lam$, which contains the unknowns $\omnO_{k}$ and $S^{(0)\prime}_{k}$. Next we insist that $S_{k}'$ is regular at $r=3$, to fix $\omnO_{k}$. Then we rearrange to find $S^{(0) \prime}_{k}$, and differentiate once to obtain $S^{(0) \prime\prime}_{k}$ which features at the next order. If we wish to construct the wavefunction we also integrate $S^{(0) \prime}_{k}$ to obtain $S^{(0)}_{k}(r)$ (whereas if we only require the frequency expansion this step may be skipped). Finally we move to the next order, sending $k \to k+1$, and repeat the steps. It is straightforward to implement the procedure in a symbolic algebra package to extend the expansion to arbitrary order. The lowest-order coefficients for perturbations of spin $|s|=0, 1, 2$ of the Schwarzschild spacetime are given in Table \ref{table-n0}. The expansion may be carried to 15th order and beyond with relative ease. For instance, the frequency of the $n=0$ gravitational ($s=-2$) QNM is 
\begin{eqnarray}
&& \sqrt{27} \omega_{l0} = \lam
-0.5 i
-1.30092593 \lam^{-1}
+0.20460391i \lam^{-2} 
-0.56376775 \lam^{-3} \nn \\ &&
+0.25454392i \lam^{-4}
-0.19978348 \lam^{-5} 
-0.05688153i \lam^{-6} 
+0.14611217 \lam^{-7} \nn \\ &&
-1.16950214i \lam^{-8} 
+0.04477903 \lam^{-9}
-3.49565902 i \lam^{-10}
-1.50293894 \lam^{-11} \nn \\ &&
-6.56185182 i \lam^{-12}
-4.40093979 \lam^{-13} + \mathcal{O}(\lam^{-14})     \label{om-n0}
\end{eqnarray}
(expressing rational coefficients to 8 decimal places). 
We note the following: (i) the method provides a neat expansion in inverse powers of $L = l+1/2$, and a globally-valid (in $r$) expansion (\ref{v-expansion}) of the QNM function; (ii) odd (even) powers of $L$ in the Schwarzschild QNM frequency expansion (\ref{omega-expansion}) have coefficients which are purely real (imaginary); (iii) the series is probably not convergent for low $L \lesssim 1$, rather one expects that Eq.~(\ref{om-n0}) is an asymptotic expansion. That is, for a given $L$, series (\ref{om-n0}) may be formally divergent; nevertheless  excellent approximations to the frequencies may be obtained by truncating at finite order $k_{m}$ such that $|\omnO_{k_{m}} L^{-k_m}| \lesssim |\omnO_{k} L^{-k}| , \; \forall k $. %The `optimal' truncation order is found from the minimum term in the series. 

\begin{table} 
\begin{tabular}{l | l | l | l | l | l | l | l | l}
     & $L^1$ & $L^0$ & $L^{-1}$ & $L^{-2}$ & $L^{-3}$ & $L^{-4}$ & $L^{-5}$ & $L^{-6}$ \\
\hline
\text{scalar,} \quad & $1$ & $-i / 2$ & $\phantom{-} \frac{7}{216}$ & $- \frac{137}{7776}i$ \, & $\phantom{-}  \frac{2615}{1259712}$ & $ \frac{590983}{362797056}i$ 
& $-\frac{42573661}{39182082048}$ & $\phantom{-} \frac{11084613257}{8463329722368} i$
 \\
\text{EM,} \quad & $1$ & $-i / 2$ & $- \frac{65}{216}$ & $\phantom{-}  \frac{295}{7776}i$ & $- \frac{35617}{1259712}$ & $ \frac{3374791}{362797056}i$ 
& $- \frac{342889693}{39182082048}$ & $\phantom{-} \frac{74076561065}{8463329722368} i$ 
 \\
\text{Grav.,} \quad & $1$ & $-i / 2$ & $-\frac{281}{216}$ & $\phantom{-}  \frac{1591}{7776}i$ & $-\frac{710185}{1259712}$ & $ \frac{92347783}{362797056}i$ & $-\frac{7827932509}{39182082048}$ & $- \frac{481407154423}{8463329722368}i$
\end{tabular}
\caption{Series expansion coefficients $w_k = \sqrt{27} \omnO_k$ for the fundamental ($n=0$) Schw.~QNM frequencies, in expansion $\sqrt{27} M \omega_{l,n=0} = \sum_{k=-1}^\infty w_k L^{-k} $, for scalar ($\beta = 1$), electromagnetic ($\beta = 0$) and gravitational ($\beta = -3$) modes.  }
\label{table-n0}
\end{table}

The method can be extended higher overtones with the ansatz
\beq
\fl v_{ln}(r) = \left[ \left(1-\frac{3}{r}\right)^n + \sum_{i = 1}^{n} \sum_{j=1}^\infty a^{(n)}_{ij} L^{-j} \left(1-\frac{3}{r}\right)^{n-i} \right] \exp \left( S^{(n)}_0(r) + L^{-1} S^{(n)}_1(r) + \ldots \right)   \label{v-higher-n}
\eeq
%
%\begin{align*}
%&\left[(r-3)^n + (a_{11}L^{-1} + a_{12}L^{-2} + \dots) (r-3)^{n-1} +
%(a_{21}L^{-1} + a_{22}L^{-2} + \dots) (r-3)^{n-2} + \dots \right.\\
%&\qquad \left. +
%(a_{n1}L^{-1} + a_{n2}L^{-2} + \dots)\right]\exp\left( S_0(r) + S_1(r)L^{-1}+S_2(r)L^{-2} +
%\dots\right) .
%\end{align*}
%
At the $i$th step of the interative procedure we require continuity of the first $n-1$ non-vanishing derivatives at $r=3$ to determine the $a^{(n)}_{ij}$, and $n$th to determine the correction to the frequency, $\omco^{(n)}_{i}$. When these conditions are imposed, we are left with an explicit equation for $S^{(n)\prime}_i(r)$.
%Note that the $n > 0$ wavefunction develops a zero at $r=3$ as $L \rightarrow \infty$.

\subsection{Results: QNM Frequency Expansion}
Expansion coefficients for the frequency of the `fundamental' mode are listed in Table \ref{table-n0}, for perturbations of spin $0$ (scalar), $1$ (electromagnetic) and $2$ (gravitational). 
Expansion coefficients for the $n=1, 2, 3$ quasinormal mode frequencies of the gravitational field on the Schwarzschild spacetime are given in Table \ref{table-n-higher}. 

%\begin{table} 
%\begin{tabular}{l | l | l | l | l | l | l}
% & $\lam^1$ & $\lam^0$  & $\lam^{-1}$ & $\lam^{-2}$ & $\lam^{-3}$ & $\lam^{-4}$ \\
%\hline
%\text{scalar,} $\beta=1$ \quad & $1$ & $-i \frac{3}{2}$ & $- \frac{53}{216}$ & $-i \frac{607}{2592}$ & $\frac{82111}{629856}$ & $i \frac{6073213}{120932352}$ \\
%\text{EM,} $\beta = 0$ \quad & $1$ & $-i \frac{3}{2}$ & $-\frac{125}{216}$ & $-i \frac{175}{2592} $ & $-\frac{3101}{629856}$ & $i \frac{303421}{120932352}$ \\
%\text{Grav.,} $\beta = -3$ \quad & $1$ & $-i \frac{3}{2}$ & $-\frac{341}{216}$ & $i \frac{1121}{2592}$ & $-\frac{538673}{629856}$ & $i \frac{63615613}{120932352}$ 
%\end{tabular}
%\caption{Series expansion coefficients $a_k$ for the first overtone ($n=1$) QNM frequencies $\sqrt{27} M \omega_{n=1} = \sum_{k=0}^\infty a_k \lam^{1-k} $.  }
%\label{table-n1}
%\end{table}

\begin{table} 
\begin{tabular}{l | l | l | l | l | l | l | l | l}
 & $\lam^1$ & $\lam^0$  & $\lam^{-1}$ & $\lam^{-2}$ & $\lam^{-3}$ & $\lam^{-4}$ & $L^{-5}$ & $L^{-6}$ \\
\hline
$n = 1$ \quad \quad & $1$ & $- \frac{3}{2}i$ & $-\frac{341}{216}$ & $\phantom{-} \frac{1121}{2592}i$ & $-\frac{538673}{629856}$ & $\frac{63615613}{120932352}i$
& $- \frac{23221986259}{39182082048}$ & $-\frac{453971786399}{940369969152} i$
 \\
$n = 2$ \quad & $1$ & $- \frac{5}{2}i$ & $-\frac{461}{216}$ & $\phantom{-} \frac{905}{7776}i$ & $-\frac{146381}{157464}$ & $\frac{115304885}{362797056}i$ 
& $-\frac{19631537719}{39182082048}$ & $-\frac{10658412628805}{8463329722368} i$ 
 \\
$n = 3$ \quad & $1$ & $- \frac{7}{2}i$ & $-\frac{641}{216}$ & $- \frac{8603}{7776}i$ & $\phantom{-} \frac{289949}{1259712}$ & $\frac{31522981}{362797056}i$ 
& $\phantom{-}\frac{125344410311}{39182082048}$ & $\phantom{-}\frac{49100100676571}{8463329722368} i$ 
\end{tabular}
\caption{Series expansion coefficients $w_k = \sqrt{27} \omcon_k$ for gravitational ($s= -2$) Schw.~QNMs of higher overtones ($n=1, 2, 3$) in expansion $\sqrt{27} \omega_{ln} = \sum_{k={-1}}^\infty w_k L^{-k} $.  }
\label{table-n-higher}
\end{table}

The lowest expansion coefficients for arbitrary spin $\beta = 1-s^2$ and arbitrary overtone number $n$ can be written as
\begin{eqnarray}
\fl \quad \sqrt{27} \, \omcon_{-1} &=& 1  \label{omco-0}  \\
\fl \quad \sqrt{27} \, \omcon_0 &=& -i N \\
\fl \quad \sqrt{27} \, \omcon_1 &=& \frac{\beta}{3} - \frac{5 N^2}{36} - \frac{115}{432} \\
\fl \quad \sqrt{27} \, \omcon_2 &=& -i N \left[ \frac{\beta}{9} + \frac{235 N^2}{3888} - \frac{1415}{15552}  \right]\\
\fl \quad \sqrt{27} \, \omcon_3 &=& - \frac{\beta^2}{27} + \frac{204 N^2 + 211}{3888} \beta + \frac{854160N^4 - 1664760 N^2 - 776939}{40310784} \\
\fl \quad \sqrt{27} \, \omcon_4 &=& i N \left[ \frac{\beta^2}{27} + \frac{1100 N^2 -2719}{46656} \, \beta  + \frac{11273136 N^4 - 52753800 N^2 + 66480535}{2902376448}  \right]   \label{omco-5}
\end{eqnarray}
where $N = n+1/2$.

\subsection{Validation\label{sec:validation}}
 In Table \ref{table-validation} the results of the expansion in powers of $L^{-1}$ are compared against the continued-fraction results of \cite{Leaver-1985}, and the WKB results of \cite{Konoplya}, for Schwarzschild gravitational modes with $l=2,3$ and $n=0,1$. At 6th order (i.e.~including all terms in series (\ref{omega-expansion}) up to and including $\omcon_5 L^{-5}$, resulting in truncation error $\mathcal{O}(L^{-6})$), the accuracy of the new method is comparable with the WKB method. It is simple to extend our method to, e.g., 12th order to obtain greater accuracy, as we show in Table \ref{table-validation}.

\begin{table} 
\fl \quad \quad \quad
\begin{tabular}{l | l | l | l | l}
  & $l=2, n=0 $ & $l=2, n=1 $ & $l=3, n=0$ &  $l=3, n=1$ \\
  \hline 
Ctd.~Frac. &  $0.373672$ $-i0.088962$ & $0.346711$ $-i0.273915$ & $0.599444$ $- i0.092703$ & $0.582644$ $-i0.281298$ \\
  \hline 
12th order & $0.373679$ $-i0.088934$ & $0.346831$ $-i 0.273844$ & $0.599443$ $-i0.092703$ & $0.582645$ $-i0.281297$ \\ 
  \hline 
6th order & $0.373642$ $-i 0.088671$ & $0.347895$ $-i0.272766$ & $0.599439$ $-i0.092684$ & $0.582713$ $-i0.281206$ \\    
% checked
  \hline 
WKB (6th)& $0.3736\phantom{00}$ $-i 0.0890$ & $0.3463\phantom{00}$ $-i0.2735$ & $0.5994\phantom{00}$ $-i0.0927$ & $0.5826\phantom{00}$ $-i 0.2813$ \\
\hline
\end{tabular}
\caption{\emph{Numerical Accuracy of the Gravitational QNM frequencies}. Comparing the results of the new expansion method at 6th and 12th order (with truncation errors of order $\mathcal{O}(L^{-6})$ and $\mathcal{O}(L^{-12})$, respectively) with numerical results from the continued fraction method \cite{Leaver-1985} and 6th order WKB results \cite{Konoplya}.}
\label{table-validation}
\end{table}

We checked the series expansion coefficients (\ref{omco-0}--\ref{omco-5}) for general spin $\beta = 1-s^2$ and overtone $n$ against the WKB results given in Eq.~(3.1) of \cite{Iyer-1987}. We find agreement up to and including coefficient $\omcon_2$ (power $L^{-2}$). This is expected, because the truncation error in the WKB results of \cite{Iyer-1987} is of order $\mathcal{O}(L^{-3})$. 

\begin{figure}
\begin{center}
\includegraphics[width=12cm]{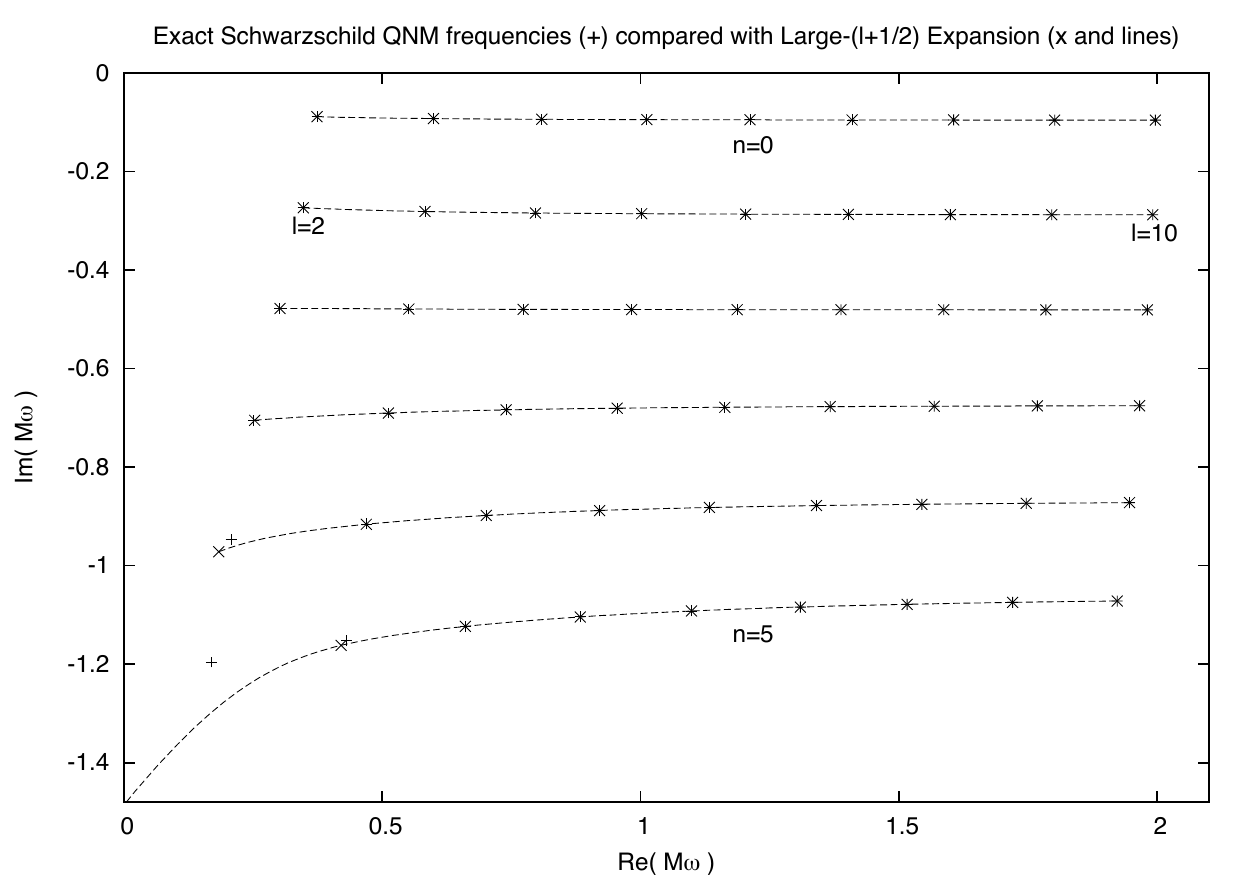}
\end{center}
\caption[]{\emph{Gravitational QNM frequencies of Schwarzschild black hole}. The exact QNM frequencies ($+$ symbol, determined via the continued-fraction method \cite{Leaver-1985}) are compared with the results of the expansion in $L^{-1}$ ($\times$ symbols, dotted lines) taken to order $\mathcal{O}(L^{-10})$.}
\label{fig-ae-curves}
\end{figure}

\begin{figure}
\begin{center}
\includegraphics[width=15cm]{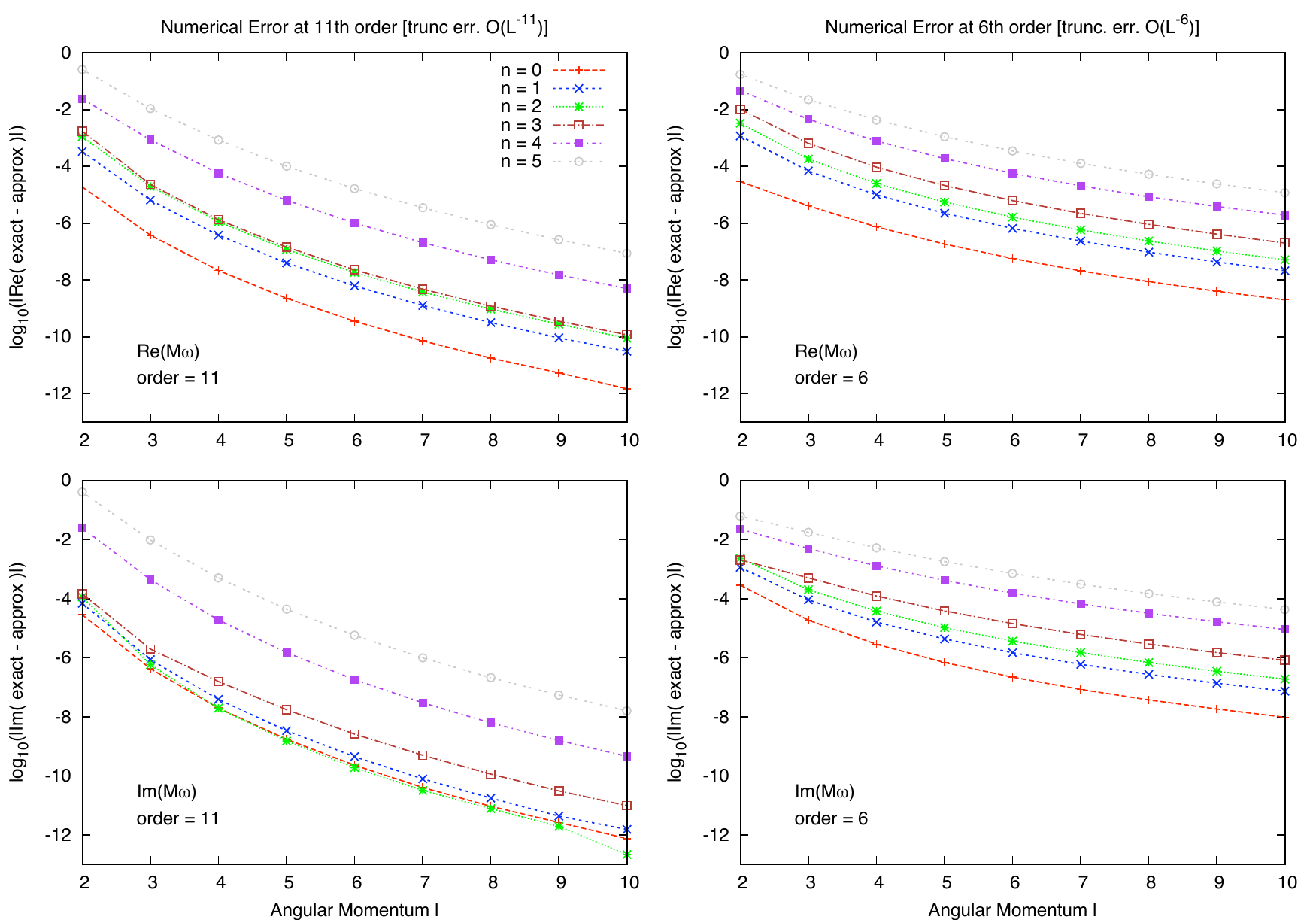}
\end{center}
\caption[]{\emph{Accuracy of Expansion of Schwarzschild gravitational QNM frequencies in inverse powers of $L=l+1/2$}. The four plots show the difference between the exact QNM frequencies (continued-fraction method) and the approximations from the asymptotic expansion method. The frequency expansion $\omega_{ln} = \sum_{k=-1}^{k_m} \omcon_k L^{-k}$ is truncated at orders $k_{m} = 5$ (left) and $k_m = 10$ (right). The upper (lower) plots show the accuracy of the real (imaginary) part of the frequency on a logarithmic scale, for overtones $n=0\ldots 5$ (dotted lines).}
\label{fig-ae-err}
\end{figure}

The accuracy of the expansion is examined in Fig.~\ref{fig-ae-curves} and \ref{fig-ae-err}. Figure~\ref{fig-ae-curves} compares the QNM frequencies found via the expansion method with the results of the continued fraction method \cite{Leaver-1985}. As expected, we find good agreement if $l \gtrsim n$, but not always in the regime $l \lesssim n$. In general, the expansions of the higher-$n$ modes are less accurate. This is clearly demonstrated in Fig.~\ref{fig-ae-err}, which shows the difference between the series expansion (taken to orders $L^{-5}$ and $L^{-10}$) and the exact frequencies, for the modes $n=0\ldots 5$.

\begin{figure}
\begin{center}
\includegraphics[width=10cm]{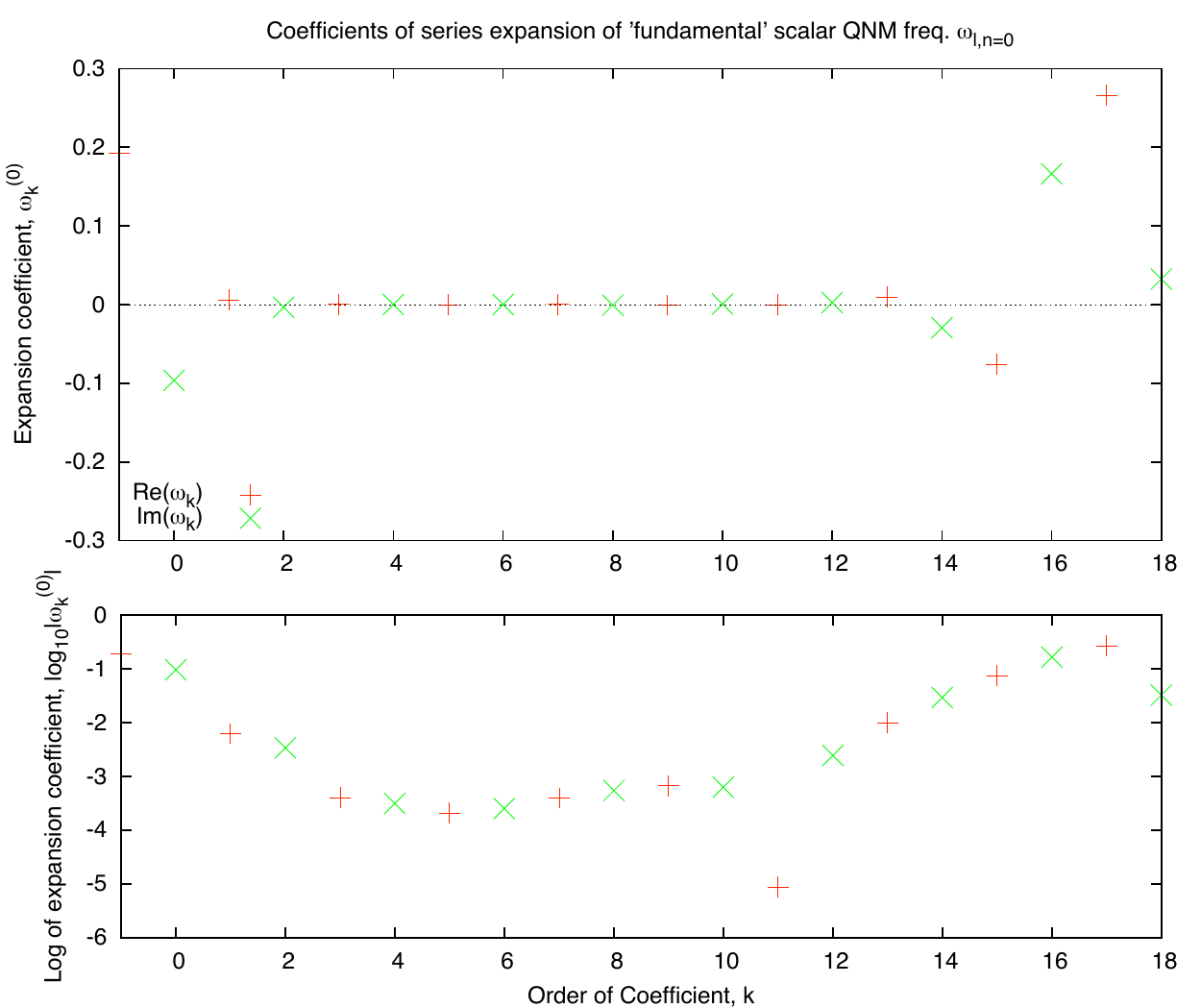}
\end{center}
\caption[]{\emph{Coefficients in the Expansion of the Fundamental Schwarzschild QNM Frequency (Scalar Field)}. The plot shows the coefficients $\omnO$ in the expansion $\omega_{l0} = \sum_{k=-1}^\infty \omnO_k L^{-k}$. The lower plot shows the same data as the upper plot, using a logarithmic scale. [N.B. The coefficients are known as rationals (i.e. exact)]. }
\label{fig-coeffs}
\end{figure}

Figure \ref{fig-coeffs} shows the magnitude of the coefficients in the series expansion of the frequencies. It shows that the coefficients $a_k$ do not seem to tend to zero as $k \rightarrow 0$, though they do become small at certain intermediate orders. This supports the view that Eq.~(\ref{omega-expansion}) is an asymptotic expansion. %In other words, a formally divergent series which may nevertheless gives excellent approximations when truncated at the appropriate order.

\subsection{Results: Outgoing coefficient $\Aout$ and the wavefunction}
A key advantage that the expansion method holds over many other approaches (see Sec.~\ref{sec-existing-methods}) is that it furnishes us with simple approximations for the wavefunctions [Eq.~(\ref{ansatz}) and (\ref{v-higher-n})] which are valid at all $r > 2$. Here we present the results for the outgoing coefficient $\Aout$ in a compact form, and examine the shape of the wavefunction. 

The integral appearing in ansatz (\ref{ansatz}) can be evaluated without much difficulty. Let us define
\begin{eqnarray}
\mathcal{X}(r) &=&  \exp  \left( i \omega \int_3^r \left[ (1+6/r)^{1/2} (1-3/r) \right] f^{-1} dr \right) \\
&=& (2 + \sqrt{3})^{- 6 i \omega} \,  \left( \frac{1+x}{2-x} \right)^{4 i \omega} \exp \left( i \omega \left[  \rstar - r(1-x)  -  \sqrt{27}  \right] \right)
\end{eqnarray}
where
\beq
x = \left( 1 + \frac{6}{r} \right)^{1/2}.
\eeq

% \frac{1}{4} (2-\sqrt{3})^6 \, e^{3-\sqrt{27}} e^{-r(1-x)} \left( \frac{1+x}{2-x} \right)^4 ,

The outgoing coefficient $\Aoutln$ defined by Eq.~(\ref{uin-def}) may be computed via
\beq
\fl \quad \quad \Aoutln = \left( - 2 \right)^{n} \left( \frac{1 + \sum_{k=1}^n \sum_{j=1}^\infty a_{kj}^{(n)} L^{-j} }{ 1 + \sum_{k=1}^n \sum_{j=1}^\infty (-2)^k a_{kj}^{(n)} L^{-j} } \right) \frac{C_{\text{out}}}{C_{\text{in}}} \exp\left( \sum_{k=0}^\infty \gamma_k^{(n)} L^{-k} \right)   \label{Aout}
\eeq
where we have defined the quantities 
\begin{eqnarray} 
C_{\text{in}} &\equiv&  \lim_{r \rightarrow 2}  \mathcal{X}(r) / e^{-i \omega \rstar} =  \exp \left( i \omega [6 - \sqrt{27} + 8\ln2 - 6 \ln (2 + \sqrt{3}) ] \right) \\
C_{\text{out}} &\equiv& \lim_{r \rightarrow \infty} \mathcal{X}(r) / e^{+ i \omega \rstar} = \exp \left( i \omega [3 - \sqrt{27} + 4 \ln 2 - 6 \ln(2 + \sqrt{3}) ] \right) \\
\gamma_k^{(n)} &\equiv& \lim_{r \rightarrow \infty}  S_k^{(n)}(r) .   \label{alp-bet-def}
\end{eqnarray}
and we choose the constant of integration such that $\lim_{r \rightarrow 2}  S_k^{(n)}(r)  = 0$. 
The leading-order function $\Sn_0(r)$ is independent of the spin of the perturbing field, and may be written compactly as
\beq
\fl \quad \quad \Sn_0 (r) = \frac{1}{2} \ln ( 2 / x ) + 2 (n+1/2) \ln \left( \frac{2 + \sqrt{3}}{x + \sqrt{3}} \right)  .  \label{S0-def1}
\eeq
Note $\Sn_0(r)$ is purely real, hence it tells us about the amplitude of the QNM in the large-$\lam$ limit. The leading-order coefficient defined in (\ref{alp-bet-def}) is therefore
\beq
\gamma^{(n)}_0 = (1/2) \ln 2 + (n+1/2) \ln \left( [2 + \sqrt{3}] / 2 \right) .
\eeq
Hence, at leading order in $L$, the outgoing coefficient is simply
\beq
\Aout \approx (-2)^n \frac{\Cout}{\Cin} \exp \left( \gamma_0^{(n)} \right)  =  (-1)^{n}
\left( 2 + \sqrt{3} \right)^{n+1/2} e^{-i \omega_{ln} (3+4\ln2)}  \label{Aout-n-gen}
% checked
\eeq
for all spins.

\begin{figure}
\begin{center}
\includegraphics[width=10cm]{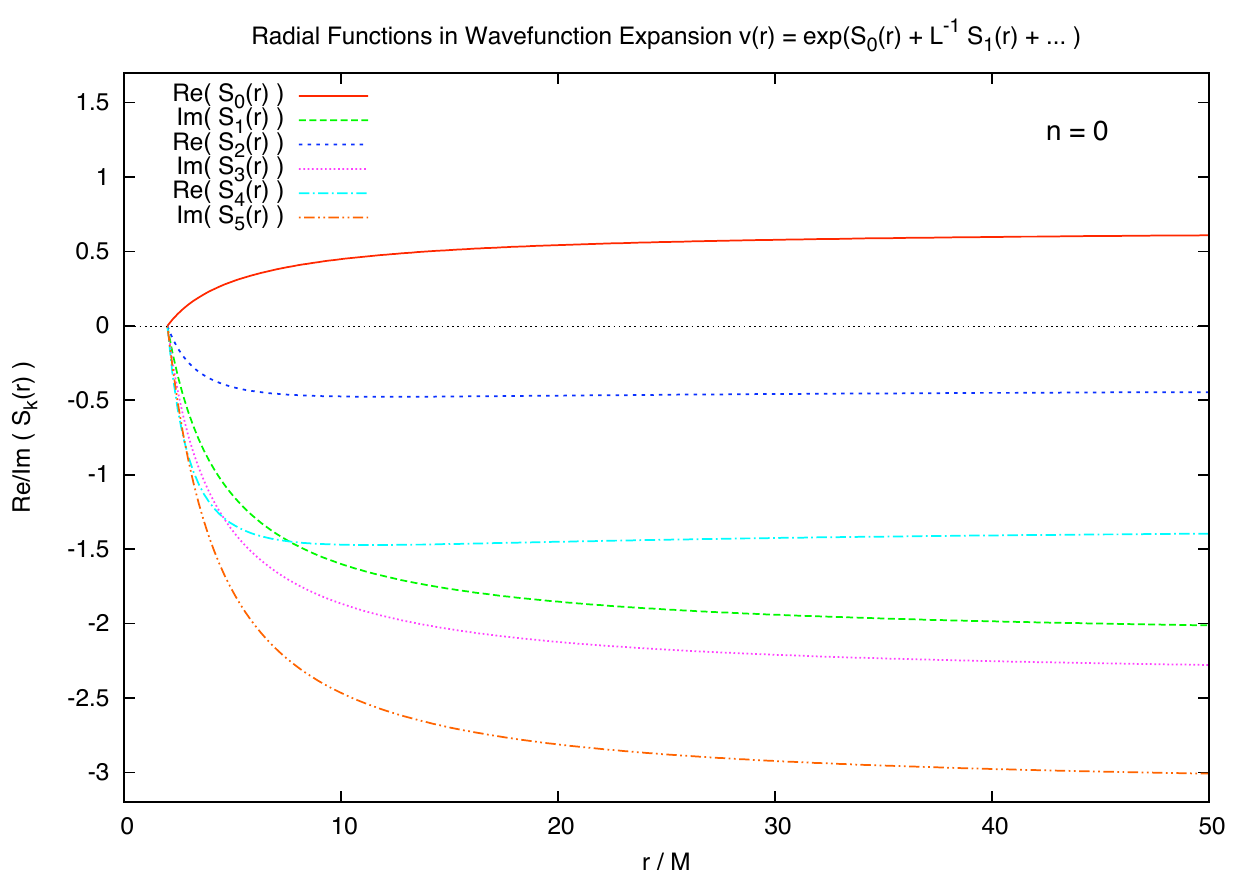}
\end{center}
\caption[]{\emph{Radial profile of the functions $\SnO_k(r)$}. Note that functions with $k$ even (odd) are purely real (imaginary). The functions for higher overtone $n$ are similar in nature.}
\label{fig-Sfn}
\end{figure}

Calculation of (\ref{Aout}) is straightforward for the `fundamental' mode, for which we find
\beq
A^{(\text{out})}_{l, n=0} = (2+\sqrt{3})^{1/2} e^{ -i \omega_{ln} (3 + 4\ln 2)} \exp \left( \sum_{k=1}^\infty \gamma_k^{(0)} L^{-k} \right)   \label{Aout-n0}
\eeq
where the first few coefficients for the fundamental mode are
\begin{eqnarray}
\sqrt{27} \, \gamma^{(0)}_1 &=&  i \left[   3 \beta - \frac{97}{48}   \right] \\
\sqrt{27} \, \gamma^{(0)}_2 &=&  \phantom{i}  \left[ \frac{\beta}{2} - \frac{577}{864} \right] \\
\sqrt{27} \, \gamma^{(0)}_3 &=&  i \left[ - \beta^2 + \frac{457}{432} \beta - \frac{1013761}{5598720} \right] \\
\sqrt{27} \, \gamma^{(0)}_4 &=&  \phantom{i} \left[  - \frac{\beta^2}{2} + \frac{20}{27} \beta - \frac{2788571}{12597120}  \right] .
\end{eqnarray}
Higher-order approximations can be found by computing the functions $\Sn_k(r)$ and the coefficients $a^{(n)}_{kj}$. Typical radial profiles of the functions $\SnO_k(r)$ are shown in Fig.~\ref{fig-Sfn}. We see that the functions start from zero at $r=2$ and tend towards a constant as $r \rightarrow \infty$, without any significant oscillation between these limits.

\begin{figure}
\begin{center}
\includegraphics[width=8cm]{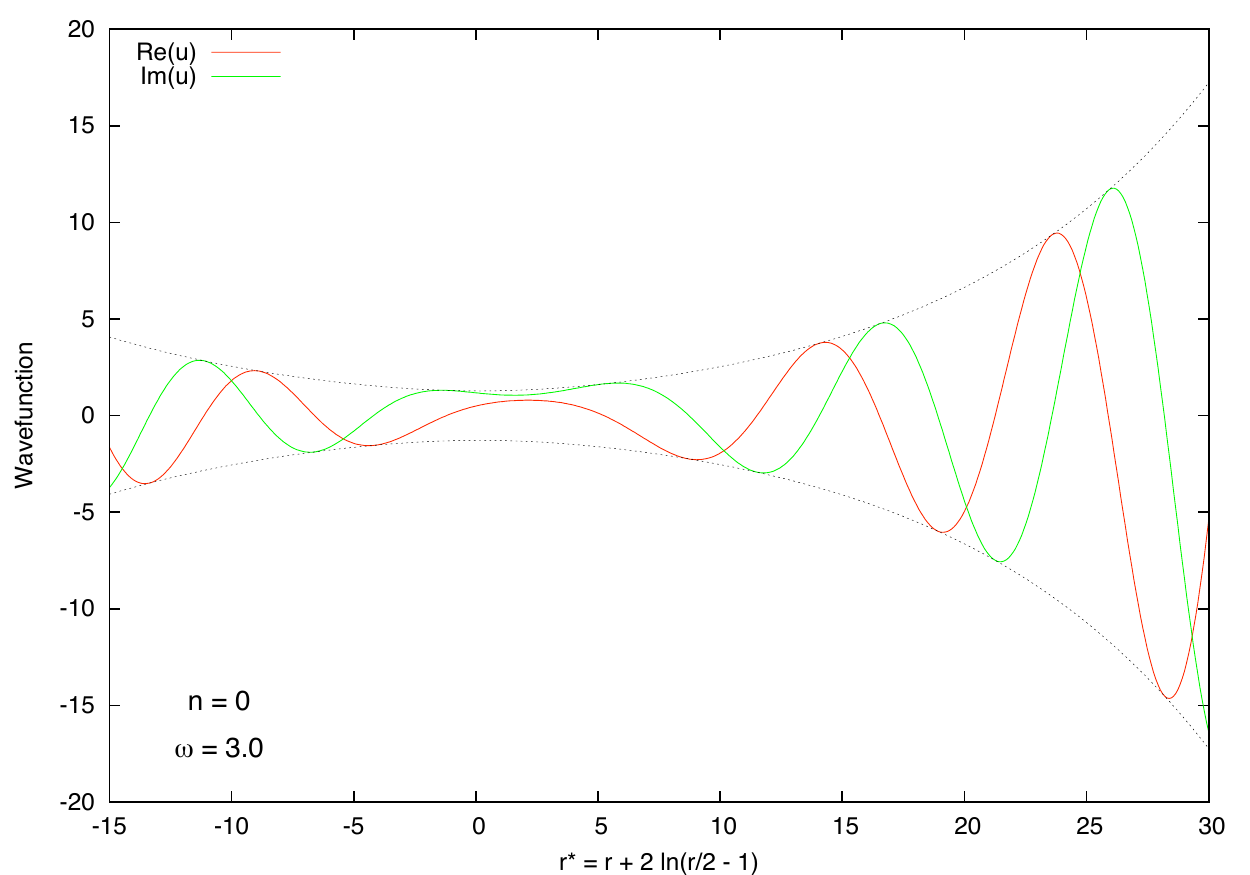}
\includegraphics[width=8cm]{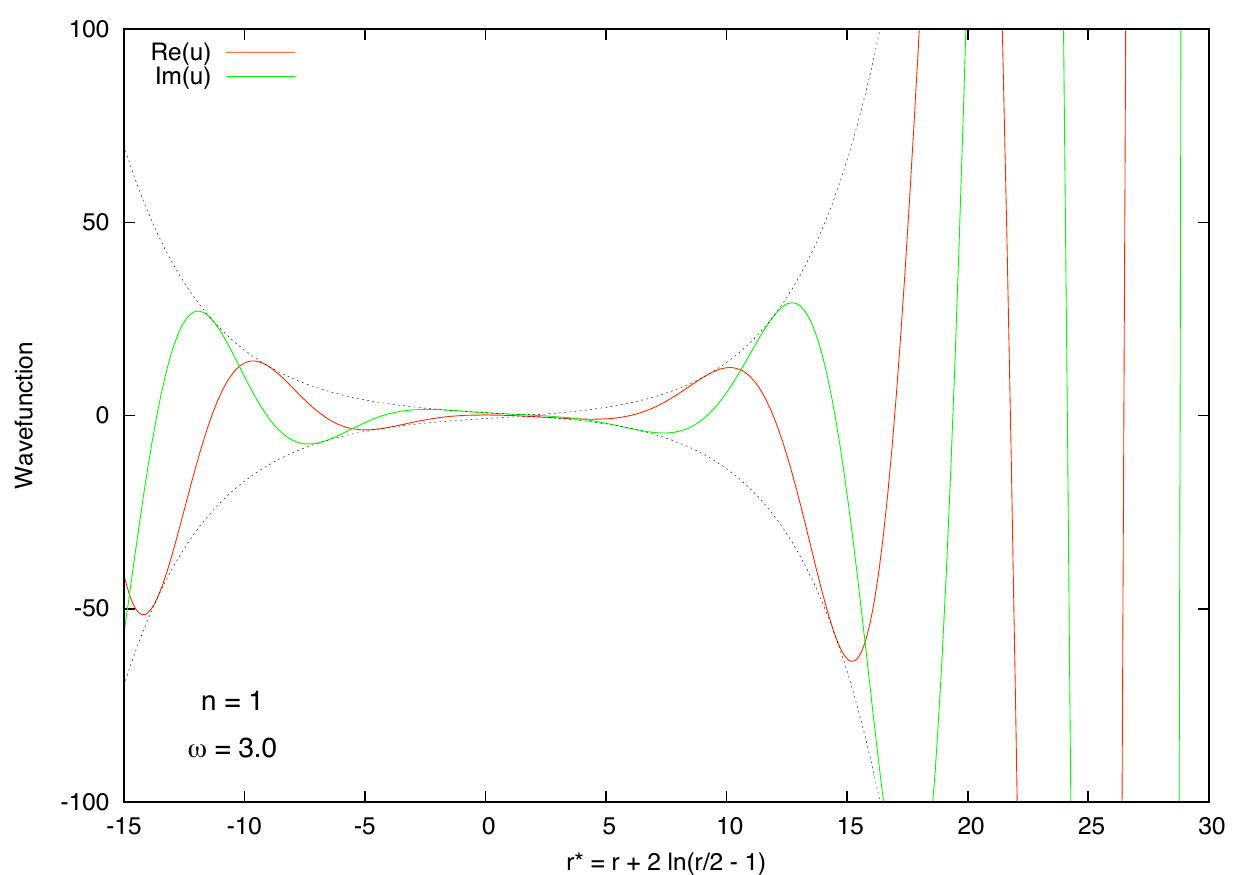}
\end{center}
\caption[]{\emph{Gravitational QNM wavefunctions of lowest overtones ($n=0, 1$) at $L = l+1/2 = 4$}. Note that the wavefunctions diverge exponentially in both limits, $\rstar \rightarrow \pm \infty$.}
\label{fig-qnm-wf}
\end{figure}

Figure \ref{fig-qnm-wf} shows examples of the QNM wavefunctions. Since the frequency has a negative imaginary part, it is no surprise to find that the wavefunctions diverge as $r \rightarrow 2$ and as $r \rightarrow \infty$. This makes it rather hard to gain physical insight from examining the radial functions alone. To assess the QNM contribution to physically-realistic scenarios, we need to compute the `QNM excitation coefficient' \cite{Berti-Cardoso-2006}. This quantity is obtained from the residues of the poles in the Green function at QNM frequency. In a forthcoming work \cite{Dolan-Ottewill-ef} we will show how to calculate the residues of the poles, and in particular the key quantity $\left. \partial \Ain / \partial \omega \right|_{\omega=\omega_{ln}}$, by combining the expansion method with standard WKB techniques. 

An alternative set of physically-relevant wavefunctions that do not diverge in the limits $r \rightarrow 2$ and $r \rightarrow \infty$ are the so-called Regge poles (shown in Fig.~\ref{fig-regge-wf}), to which we now turn our attention.

\section{The Expansion of Schwarzschild Regge Poles\label{sec-regge}}
For a given angular momentum $l$, there are an infinite set of QNMs with frequencies $\omega_{ln}$ defined by $A^{\text{(in)}}_{l \omega_{l n}} = 0$. Conversely, for a given frequency $\omega$ there are an infinite set of Regge poles, corresponding to complex angular momenta $\lambda_{\omega n} = l_{\omega n} + 1/2$, defined by the condition $A^{\text{(in)}}_{l_{\omega n} , \omega} = 0$. 
Regge poles are a key concept in the so-called `Complex Angular Momentum' (CAM) method \cite{CAM}, which has been applied in a variety of contexts. For example, Regge poles received much attention in high-energy physics in the 1960s \cite{Collins}. In the context of black hole physics, it has been shown that the scattering amplitude for a monochromatic planar wave impinging upon a black hole can be split into a background integral and a sum over Regge poles \cite{Andersson-Thylwe-1994, Andersson-1994}. The background integral (approximately) corresponds to the `classical' scattering cross section, and the Regge poles correspond to regular oscillations in intensity-vs-angle plot caused by the interference of parts of the wavefront travelling close to rays passing in opposite senses around the black hole (see \cite{Andersson-Jensen} for a short review).

Many of the methods used for finding QNMs can also be used to find Regge poles \cite{Decanini-Folacci-Jensen, Decanini-Folacci-2009}. It is perhaps no surprise to find it is straightforward to expand Regge poles $\lambda_{\omega n}$ in inverse powers of $\omega$. We simply insert the expansion
\begin{eqnarray}
\fl \lambda_{\omega n} &=&  \lamcon_{-1} \omega +  \lamcon_0 +  \lamcon_1 \omega^{-1} +  \lamcon_2 \omega^{-2}  + \ldots  \label{rp-expansion} \\
\fl v_{\omega n}(r) &=& \left[ \left(1-\frac{3}{r}\right)^n + \sum_{i = 1}^{n} \sum_{j=1}^\infty b^{(n)}_{ij} \omega^{-j} \left(1-\frac{3}{r}\right)^{n-i} \right] \exp \left( \Tn_0(r) + \omega^{-1} \Tn_1(r) + \ldots \right)
\end{eqnarray}
into the differential equation (\ref{rad-eq-2}), and group like powers of $\omega$, to obtain a set of equations.

We then solve order-by-order to determine the coefficients $\lamcon_k$, $b^{(n)}_{ij} $ and radial functions $\Tn_k$, as before. We find that the first few expansion coefficients are
\begin{eqnarray}
\fl \lambda_{-1}^{(n)} &=& \sqrt{27}  \label{lamco-0} \\
\fl \lambda_0^{(n)} &=& i N \\
\fl \lambda_1^{(n)} &=& \frac{60 N^2 - 144 \beta + 115}{432 \sqrt{27}}   %checked 
 \\
\fl \lambda_2^{(n)} &=& - i N \, \left(  \frac{1220N^2 - 6912\beta + 5555}{419904}  \right)  %checked 
\\
\fl \lambda_3^{(n)} &=& -\frac{18 \beta^2}{6561 \sqrt{27}} + \frac{[2316N^2 + 479] \beta}{104976 \sqrt{27}} -  \frac{2357520 N^4 + 19382280 N^2 + 2079661}{1088391168 \sqrt{27}} %checked
\\
\fl \lambda_4^{(n)} &=& iN \left[ \frac{8 \beta^2}{19683} - \frac{5 \beta [3716N^2 + 2291]}{17006112}  + \frac{144920784N^4 + 1871793480 N^2 + 593617841}{2115832430592} \right]  % checked 
\label{lamco-5}
\end{eqnarray}
where $N = n+1/2$ and $\beta = 1-s^2$. 
%Further results are given in Table ...  
We have verified that the expansion in inverse powers of $\omega$ given in Eq.~(\ref{lamco-0}--\ref{lamco-5}) is fully consistent with the WKB result recently obtained by D\'ecanini and Folacci (\cite{Decanini-Folacci-2009}, equation~(12) and (16)), up to and including order $\omega^{-2}$ (but not beyond). As in the QNM case (c.f. Sec.~\ref{sec:validation}) this is consistent with the expected truncation error in the WKB results. We may take the expansion to higher orders if desired; for example, the `fundamental' $n = 0$ Regge pole for the gravitational perturbation is 
\begin{eqnarray}
\fl \lambda_{\omega, n=0} &\approx& \sqrt{27}\omega + \frac{i}{2} + \frac{281 \sqrt{3}}{1944 \omega} - \frac{6649i}{209952 \omega^2} - \frac{1044601\sqrt{3}}{153055008 \omega^3}  + \frac{926224193 i}{264479053824 \omega^4}   \nn \\
\fl  && 
- \frac{184851431845 \sqrt{3}}{257073640316928 \omega^5} 
-\frac{71361067332161i}{166583718925369344 \omega^6} - \frac{14390928366797903 \sqrt{3}}{161919374795459002368 \omega^7} \nn \\
\fl && + \frac{7717840397981223883i}{139898339823276578045952 \omega^8} 
+ \frac{14117571610293670714747 \sqrt{3}}{1223830676774023504745988096 \omega^9}
+ \ldots
\label{regge-n0-expansion}
\end{eqnarray}
%The coefficients in this series agree with those obtained from expanding the WKB result (\cite{Decanini-Folacci-2009}, Eq.~(12)) to order up to and including $\omega^{-2}$, but not beyond (i.e. coefficients of $\omega^{-m}, \; m > 2$). 
Note that the expansion (\ref{regge-n0-expansion}) only works in the regime $M \omega \gtrsim 0.5$, but is not appropriate or effective near $\omega \sim 0$.

In Table \ref{tbl-regge-validation} the results of the expansion method are validated against the (approximate) WKB and the (exact) continued-fraction results presented in \cite{Decanini-Folacci-2009}. We find that the high-order expansions are in general more accurate than the WKB results, at large and intermediate $\omega$. However, at small $\omega \lesssim 0.5$ the expansion (\ref{rp-expansion}) does not converge well, hence the results are of limited use. The WKB method \cite{Decanini-Folacci-2009}, by contrast, gives results which seem relatively robust down to $\omega \sim 0.25$. 

\begin{table}
\begin{tabular}{l | l l | l l | l}
 & & & & & Expected truncation \\
 & $\text{Re}(\lambda_{\omega,n=0} )$ & $\text{Im}(\lambda_{\omega,n=0} )$ & Re(error) & Im(error) & error (approx.) \\
 \hline
$M \omega = 2.5$ & & & & & \\
\quad Exact & $13.0897823$ & $0.4950209 i$ & & & \\
\quad 15th ord. & $13.089782327$ & $0.495020871 i$ & $(0.0\%)$ & $(0.00002\%)$ & $(2. \times 10^{-12} + 3. \times 10^{-13} i)$ \\ 
\quad 8th ord. & $13.089782322$ & $0.495020835 i$ & $(0.0\%)$ & $(0.00002\%)$ & $(2. \times 10^{-7} + 2. \times 10^{-6} i)$ \\ 
\quad WKB  & $13.0899530$ & $0.4950172 i$ & $(-0.0013\%)$ & $(0.0007\%)$ & \\
\hline
$M \omega = 1.0$ & & & & & \\
\quad Exact & $5.4358037$ & $0.4714533 i$ & & & \\
\quad 15th ord. & $5.4358037$ & $0.4714533 i$ & $(0.0\%)$ & $(0.0\%)$ &  $(3. \times 10^{-7} + 1. \times 10^{-7} i)$ \\
\quad 8th ord. & $5.4357860$ & $0.4714045 i$ & $(0.0003 \%)$ & $(0.01 \%)$ &  $(1.5 \times 10^{-4} + 4.3 \times 10^{-4} i)$ \\
\quad WKB & $5.4381254$ & $0.4713563 i$ & $(-0.04\%)$ & $(0.02\%)$ &  \\
\hline
$M \omega = 0.5$ & & & & & \\
\quad Exact & $3.0310943$ & $0.4112060 i$ & & & \\
\quad 15th ord. & $3.0319568$ & $0.4106489 i$ & $(-0.028\%)$ & $(0.135\%)$ &  $(1.8 \times 10^{-4} + 2.7 \times 10^{-3} i)$ \\
\quad 8th ord. & $3.0243829$ & $0.4019403 i$ & $(0.22\%)$ & $(2.25\%)$ &  $(2.0  \times 10^{-2} +  2.7 \times 10^{-2} i)$ \\
\quad WKB & $3.0434848$ & $0.4109477 i$ & $(-0.41\%)$ & $(0.06\%)$ &  \\
\end{tabular}
\caption{\emph{Numerical Validation of Regge Pole Expansion Method}. The Regge pole expansion Eq.~(\ref{rp-expansion}) is compared against the results of the continued-fraction method (`Exact') and the WKB method obtained by D\'ecanini \& Folacci \cite{Decanini-Folacci-2009}. The accuracy of the expansions of $\lambda_{\omega n}$ at orders $\omega^{-7}$ and $\omega^{-14}$ are compared. Expected `truncation errors' (found from the magnitude of the last terms in the series) are stated in the last column. }
\label{tbl-regge-validation}
\end{table}

\begin{figure}
\begin{center}
\includegraphics[width=8cm]{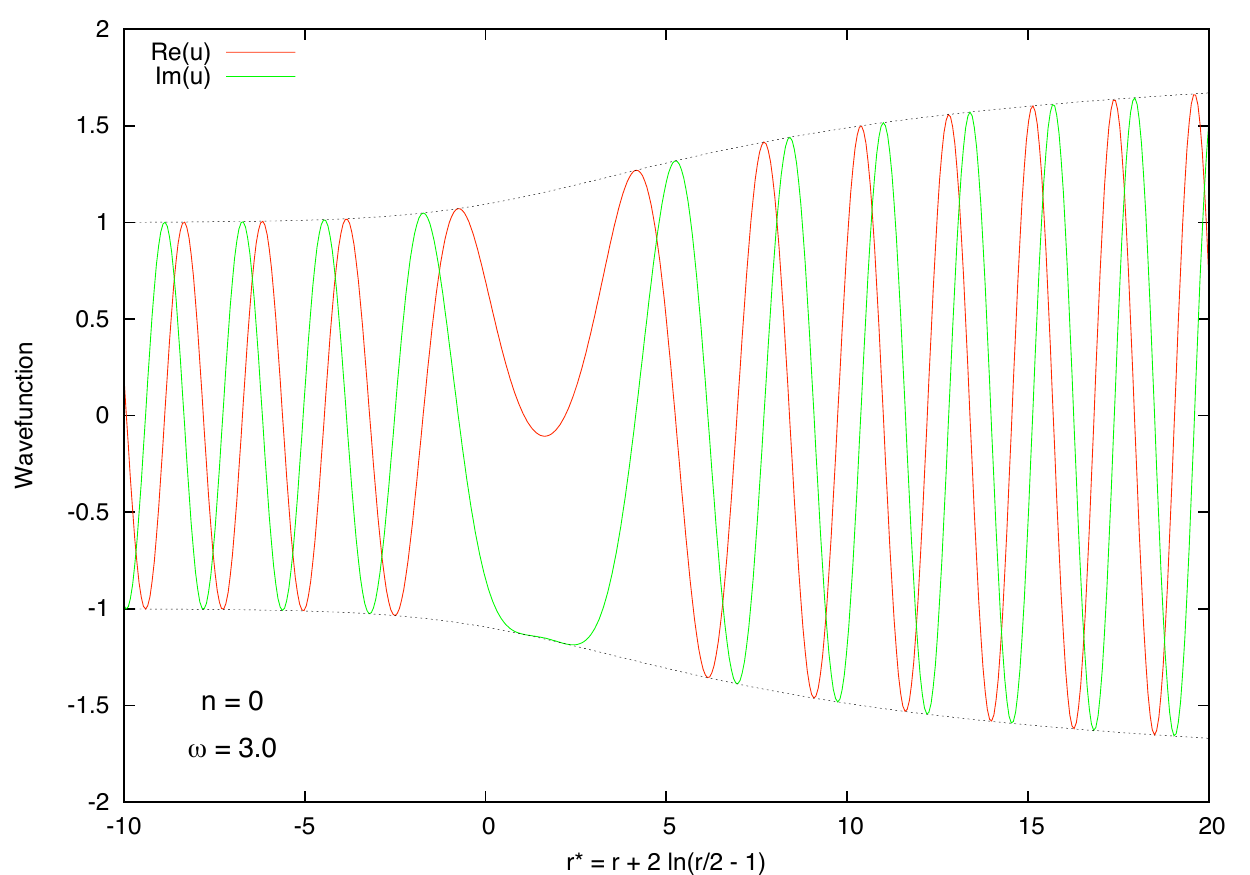}
\includegraphics[width=8cm]{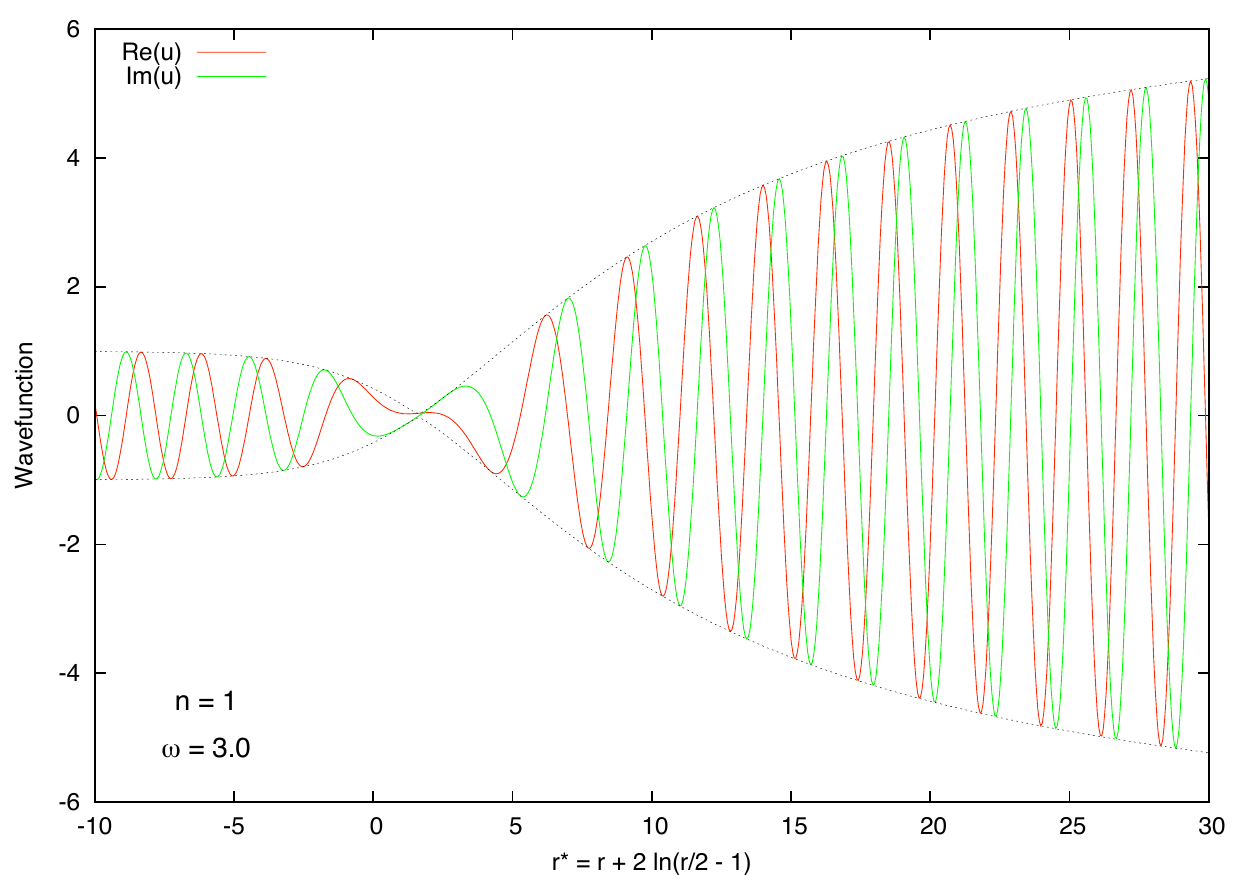}
\includegraphics[width=8cm]{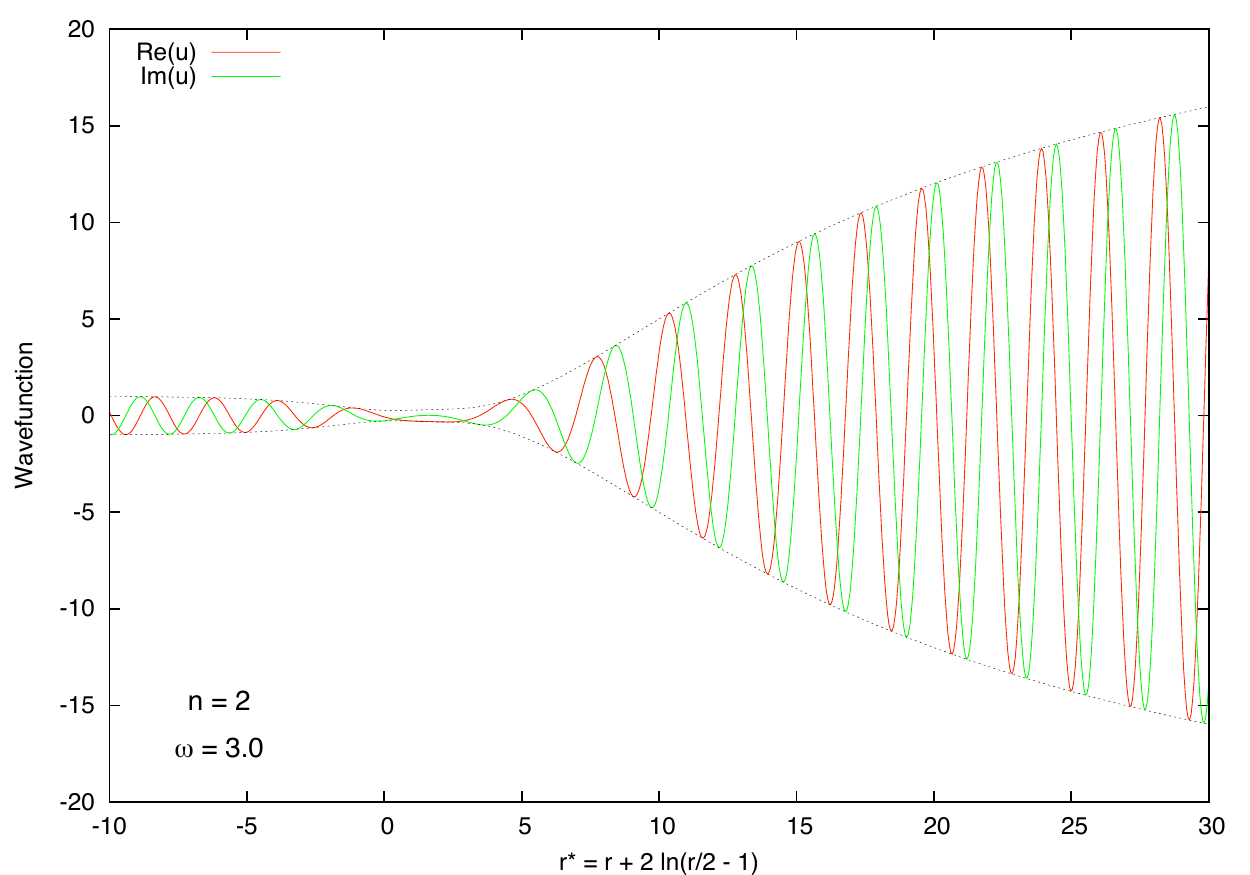}
\end{center}
\caption[]{\emph{Regge pole wavefunctions of lowest overtones ($n=0, 1, 2$) of gravitational perturbations of a Schwarzschild black hole at $M \omega = 3.0$}.  }
\label{fig-regge-wf}
\end{figure}

Wavefunctions of the lowest overtones ($n=0,1,2$) at $\omega = 3$ are shown in Fig.~\ref{fig-regge-wf}. Note that, unlike the QNM wavefunctions (\ref{fig-regge-wf}), they do not diverge in the limits $\rstar \rightarrow \pm \infty$. 

%The RP values calculated from the coefficients in Table ...  compare well with the results of the WKB method, recently given in \ref{Decanini-Folacci-2009}.

\section{Expansion of QNMs in Spherically-Symmetric Spacetimes\label{sec-ss}}

Having established the utility of the method on the Schwarzschild spacetime, 
let us now seek to generalise the analysis. In Sec.~\ref{subsec:crit-orb} and \ref{subsec:wave-eq} we derive the leading-order terms for the QNM frequencies in a general 
static spherically-symmetric spacetime. 
The analysis highlights the link between the geodesic equations and the expansion method. In Sec.~\ref{sec-examples} we look more carefully at some specific cases: %electromagnetic and gravitational perturbations of a
 Reissner-Nordstr\"om; Schwarzschild-deSitter; Nariai; higher dimensional black holes; and the canonical acoustic hole.

\subsection{Critical Orbits\label{subsec:crit-orb}}
Let us assume a $(d+2)$-dimensional static spherically-symmetric spacetime with line element
\beq
ds^2 = -f(r) dt^2 + g^{-1}(r) dr^2+ h(r) d\Omega_{(d)}^2   \label{le-ss}
\eeq
where $d\Omega_{(d)}^2$ is the natural line element on $S^d$ and we assume that $h(r) > 0$ everywhere.  This includes all the spacetimes mentioned above, 
indeed for all of them except Nariai we may take $h(r) = r^2$ 
while for Nariai $h(r) =1$.

Null geodesics on this spacetime may be taken to lie in the equatorial plane where 
the orbital equation may be written  as
\beq
\frac{f(r)}{g(r)h^2(r)} \left( \frac{d r}{d \phi} \right)^2 = \frac{1}{b^2} - \frac{f(r)}{h(r)} 
\eeq
Here the constant $b = L/E$ may be interpreted as an `impact parameter',  where $L = h(r) \dot{\phi}$ and $E = f(r) \dot{t}$ are constants of the motion 
and $\dot{\ }$ denotes differentiation with respect to an affine parameter. 

Let us define a new function 
\beq
k^2(r,b) = \frac{1}{b^2} - \frac{f(r)}{h(r)}.   \label{h2-def}
\eeq
Now assume that there is a critical impact parameter $b=b_c$, for which $k^2(r, b_c)$ has a repeated root so
 there exists some pair of constants $\{r_c, b_c\}$ such that
\beq
k^2(r_c, b_c) = 0  \quad \quad \text{and} \quad \quad \frac{\partial k^2}{\partial r} (r_c, b_c) = 0 ,
\eeq
Assuming that the repeated root is a double root let us then write
\beq
k_c(r) = \mathrm{sgn}(r-r_c) \sqrt{k^2(r,b_c)} = (r-r_c) K(r) ,  \label{kc-def}
\eeq
so $k_c(r)$ is positive for $r > r_c$ and negative for $r < r_c$.
 
It is conventional to study the stability properties of this orbit by writing the radial equation as
\beq
\dot{r}^2 = \frac{L^2 g(r)}{f(r)} k^2(r,b)  \equiv V_r(r)
\eeq
Then the period of the circular orbit is given by $2\pi/\Omega_c$ where
\beq
\Omega_c = \frac{\dot{\phi}}{\dot{t}} = \sqrt{\frac{f(r_c)}{h(r_c)}} = \sqrt{\frac{f'(r_c)}{h'(r_c)}} = 1/b_c
\eeq
and the instability timescales are determined by the Lyapunov exponent~\cite{Lyapunov2}
\beq
\lambda = \sqrt{\frac{V_r''(r_c)}{2 \dot{t}^2}} =  \sqrt{g(r_c)h(r_c)} K(r_c) \label{lyap-def} .
\eeq
% F. Pretorius and D. Khurana, Classical Quantum Gravity
%24, S83 (2007).
%\beq
%\gamma &= \sqrt{\frac{\dot{t}^2}{2 V''(r_c)}} =  \sqrt{\frac{f(r_c)}{2 g(r_c) [f(r_c) h''(r_c)  - f''(r_c) h(r_c)]}}
%\eeq

\subsection{Wave equations\label{subsec:wave-eq}}

To motivate our form of the wave equation we start by considering a scalar wave
equation with potential $(\square-V)\phi=0$.  Using the fact that the eigenvalues of the
Laplacian on $S^d$ are  $-l(l+d-1)$ (with multiplicity $\bigl({d+l \atop d}\bigr) - \bigl({d+l-2 \atop d}\bigr))$,
the wave equation separates to give a radial equation of the form
\beq
\fl \sqrt{fg} h^{-d/2} \frac{d}{dr} \left( \sqrt{fg} h^{d/2} \frac{d R}{d r} \right) +  \left[  \omega^2 - \frac{f(r)}{h(r)} \left( L^2 - (d-1)^2/4 \right) -f V(r) \right] R = 0   \label{rad-eq-general2}
\eeq
where $L = l + (d-1)/2$.  Introducing $r_*$ by $\frac{d r_*}{dr} = 1/\sqrt{fg}$ and changing 
dependent variable to $R = h^{-d/4} u$, this may be rewritten as
\beq
 \frac{d^2u}{dr_*^2} +  \left[  \omega^2 - \frac{f(r)}{h(r)}  \left(L^2 - (d-1)^2/4\right) -f  V_\mathrm{eff}(r)  \right] u = 0   \label{rad-eq-general}
\eeq   
where
\beq
V_\mathrm{eff}(r) = V(r) + h^{-d/4} f^{-1} \sqrt{fg}\frac{d}{dr} \left( \sqrt{fg}  \frac{d h^{d/4}}{d r} \right)
\eeq
To provide a framework of sufficient generality we shall now study Eq.~(\ref{rad-eq-general})
but allow $V_\mathrm{eff}(r)$ to contain $L$ dependence at lower orders than $L^2$, so
\beq
V_\mathrm{eff}(r) = V_{-1}(r) L  + V_0(r) + V_1(r) L^{-1} + \dots .
\eeq
This form is motivated by, for example, gravitational perturbations of the Reissner-Nordstr\"om spacetime which we study below.

Next we introduce the corollary to ansatz (\ref{ansatz}), namely,
\beq
u(r) = \exp \left( i \omega \int^{r_\ast} b_c k_c(r) dr_\ast \right) v(r) \ . \label{ansatz-general}
\eeq
where $k_c(r)$ was defined in Eq.~(\ref{kc-def}). 
In all the cases we shall assume that we have a horizon as $r_* \to - \infty$, so  
$f(r) \to 0$ and correspondingly $b_c k_c(r) \rightarrow -1$. 
In addition as $r_* \to \infty$ we either have an asymptotically flat region or 
a cosmological horizon, so $f(r)/h(r) \to 0$ and $b_c k_c(r) \rightarrow +1$. 
Our ansatz thus encapsulates the desired ingoing and outgoing boundary conditions 
for the wave. Note that our assumptions at this stage have excluded, for example,
 Schwarzschild-AdS space-time. 

We now substitute ansatz (\ref{ansatz-general}) in to the radial equation (\ref{rad-eq-general}) 
to obtain
\begin{eqnarray}
\fl fg v'' + \sqrt{fg} \left( (\sqrt{fg})' + 2 i b_c \omega k_c\right)v'  \nonumber \\
\lo{+} \left[(1- b_c^2 k_c^2)\omega^2 - \frac{f}{h}( L^2 - (d-1)^2/4) -f V_\mathrm{eff} + i \omega b_c \sqrt{fg} \frac{dk_c}{dr} \right] v = 0 \ .
\end{eqnarray}
Once again, we expand $\omega$ and $v(r)$ in inverse powers of $L$, using (\ref{omega-expansion}) and (\ref{v-expansion}). Making use of the key property $1-b_c^2 k_c^2(r) =  b_c^2 f(r) / h(r)$ 
we balance at order $L^2$ for all $r$ by taking
\beq
\omega_{-1} =  \frac{1}{b_c} = \Omega_c.
\eeq
Next at order $L$ we require
\beq
\fl 2i \sqrt{fg} (r-r_c)K   S_0' + 2(1-b_c^2(r-r_c)^2 K^2) \Omega_c \omega_0 - f V_{-1}
  + i \sqrt{fg} \left[K-(r-r_c) K'\right] =0 . \label{eq:gen-orderL}
\eeq
Setting $r=r_c$, requiring $S_0'$ to be regular there, we find
\beq
\omega_0 = \frac{f_c V_{-1\,c}}{2 \Omega_c} - i \frac{1}{2} \sqrt{g_c h_c} K_c
 = \frac{f_c V_{-1\,c}}{2 \Omega_c} - i \frac{1}{2} \lambda .
\eeq
where a subscript $c$ on a function of $r$ denotes that function evaluated at $r=r_c$. Note that, to leading order, the imaginary part of the frequency is set by the Lyapunov exponent (\ref{lyap-def}), as anticipated in Eq.~(\ref{leading-order}) (see also \cite{Lyapunov1,Lyapunov2,Lyapunov3}).
Substituting back into Eq.~(\ref{eq:gen-orderL}), we may write an explicit and manifestly regular equation for $S_0'$:
\beq
2i  \sqrt{fg}K  S_0' = \Delta[ f V_{-1}]
  - i \Delta[\sqrt{fg}K] - i \sqrt{fg} K' + 2 (\omega_0/\Omega_c) (r-r_c) K^2 , \label{eq:gen-orderLS0}
\eeq
where we have introduced the notation
\beq
\Delta[F](r) = \frac{F(r)-F(r_c)}{r-r_c}  \ .
\eeq
This process may readily be extended to higher orders yielding
\begin{eqnarray}
\fl \omega_{p+1} = \bigl( - f_c g_c \bigl(S''_{p\,c} 
+\sum\limits_{i=0}^{p} S'_{i\,c} S'_{p-i\,c} \bigr) - i (\omega_p/\Omega_c) \sqrt{fg}_c K_c - \sqrt{fg}_c(\sqrt{fg})'{}_c S'_{p\,c} \nonumber\\
\lo{-} \sum\limits_{i=0}^{p} \omega_i \omega_{p-i} + f V_{p\,c} - \frac{f_c (d-1)^2}{4 h_c} \delta_{p0}\bigr)/(
2 \Omega_c)
\end{eqnarray}
and
\begin{eqnarray}
\fl 2i  \sqrt{fg}K  S_{p+1}' = -\Delta\biggl[ fg \Bigl(S''_{p} +\sum\limits_{i=0}^{p} S'_{i} S'_{p-i} \Bigr)\biggr]
-\Delta\bigl[ \sqrt{fg} (\sqrt{fg})' S'_{p}\bigr]
-i (\omega_p/\Omega_c) \Delta[ \sqrt{fg} K]\nonumber\\
\lo{-} \Delta\left[\frac{f (d-1)^2}{4 h}\right] \delta_{p0}+\Delta [ f V_{p}] 
-i (\omega_p/\Omega_c) \sqrt{fg} K'\nonumber\\
\lo{-}2 i \sqrt{fg} K \sum\limits_{i=0}^{p} \omega_i S'_{p-i}/\Omega_c
+\Bigl( 2 \Omega_c \omega_{p+1} + \sum\limits_{i=0}^{p} \omega_{i} \omega_{p-i}/\Omega_c^2 \Bigr) 
  (r-r_c) K^2 .
\end{eqnarray}

\section{Examples\label{sec-examples}}

To further demonstrate the utility of the above approach, we now apply to five cases of current interest. 

%The analysis of the previous section, while general, invokes assumptions about (i) the form of the wave equation, and (ii) the boundary conditions, which are not valid for all cases of interest. However, the expansion approach is flexible and can be modified on a case-by-case basis. Here, to demonstrate its utility, we apply the method to four cases of current interest. We hope this will inspire readers to apply the method to investigate  spacetimes of particular physical interest.

%(though many other applications are possible for the interested reader): (i) gravitational perturbations of the Reissner-Nordstrom black hole; (ii) Schwarzschild-de Sitter; (iii) higher-dimensional black holes; and, (iv) the canonical acoustic hole. We hope that this may inspire some more in-depth analyses of specific situations of physical interest, in the future. 

\subsection{Reissner-Nordstr\"om Black Hole.}
The expansion method may be applied to deduce the QNM frequencies and wavefunctions of the charged (Reissner-Nordstr\"om) black hole, with a charge-to-mass ratio $q$. The relevant radial equation is \cite{Chandrasekhar, Leaver-1990, Berti, Onozawa}
\beq
f(r) \frac{d}{d r} \left(f(r) \frac{d u}{d r} \right)  + \left[ \omega^2 - f(r) \left( \frac{L^2 - 1/4}{r^2} - \frac{\kappa_{|s|}}{r^3} + \frac{q^2 \eta_{|s|}}{r^4}  \right) \right] u = 0
\eeq
where 
\beq
 \kappa_{|s|} =
 \left\{  \begin{array}{ll}
 -2 & \quad  |s| = 0 \quad \text{Scalar}  \\
 3 - \sqrt{9 + 4 q^2 (L^2 - 9/4)} & \quad  |s| = 1 \quad  \text{Electromagnetic} \\
 3 + \sqrt{9 + 4 q^2 (L^2 - 9/4)} \quad & \quad  |s| = 2  \quad \text{Gravitational} 
\end{array} \right.
\eeq
and $\eta_{|s|} = \{2, 4, 4\}$ for $|s| = \{0,1,2\}$ and $f(r) = 1 - 2/r + q^2 / r^2$. 
%Note that this radial equation is not in the form (\ref{rad-eq-general}). Nevertheless, we may still apply the expansion method. 
For $|s|=1,2$, we may expand $\kappa_{|s|}$ in inverse powers of $L$, e.g. $\kappa_{|s|} = \pm 2 q L + 3 + \mathcal{O}(L^{-1})$. The resultant terms at order $L$ in the radial equation mean that the odd (even) powers of $L$ are no longer purely real (imaginary). 

Applying the method, we find the `fundamental' ($n=0$) gravitational mode is 
\begin{eqnarray}
b_c \, \omega_{l,n = 0} &=& L - \left[ \frac{2q}{\alpha + 3} + \frac{i}{2} \left(\frac{2 \alpha}{\alpha + 3}\right)^{1/2} \right]  \nn \\  
&& -  \left( \frac{(1+\alpha)(11 - 25\alpha + 84\alpha^2)}{64(\alpha+3)\alpha^2 L} - i \, \frac{3\sqrt{2} \, q (\alpha^2-1) }{8 (3\alpha + \alpha^2)^{3/2} L} \right)   +  \mathcal{O}(L^{-2})  \label{rn-freq}  %checked
\end{eqnarray}
where $b_c = (3+\alpha)^{3/2} / (2 (\alpha+1))^{1/2}$ and $\alpha = \sqrt{9 - 8q^2}$ and $q$ is the charge-to-mass ratio of the black hole. Note that the formula breaks down, not when the twin horizons merge and disappear at $q = 1$, but rather when the unstable circular orbit disappears, at $q = 3 / \sqrt{8}$. 
More accurate numerical results for $l=2$ are given in, e.g., ~\cite{Leaver-1990, Andersson-Araujo-Schutz}. 

For the critically-charged case, $q = 1$, we obtain
\begin{eqnarray}
\omega_{l,n=0} &=& \frac{ L}{4} - \frac{2 + \sqrt{2} i}{16} - \frac{35}{256 L} - \frac{560 - 131 \sqrt{2}i}{8192 L^2} \nn \\
&&  - \frac{5031 - 2096 \sqrt{2} i}{131072 L^{3}}  - \frac{392512 - 137605\sqrt{2}i}{16777216L^4} + \mathcal{O}(L^{-5})
\end{eqnarray}
The accuracy of the eigenvalues obtained from our series is competitive with the most accurate values in the literature \cite{Onozawa}. For example, at 10th order we obtain $\omega_{4,0} = 0.9657626 - 0.08700129i$ which agrees with the $l=4$ mode presented in \cite{Onozawa} to the five significant figures given there. 

\subsection{Schwarzschild-de Sitter Black Hole.}
The expansion method may also be applied to non-asymptotically flat spacetimes. For example, in Schwarzschild-de Sitter (SdS) spacetime, the line element takes the form (\ref{le-ss}) with $f(r) = g(r) = 1 - 2/r - \Lambda r^2 / 3$ and $h(r)=r^2$, where $0 \le \Lambda \le 1/9$. The `fundamental' ($n=0$) QNMs of the scalar field in SdS are given by
\begin{eqnarray}
\fl  && \sqrt{\frac{27}{1-9\Lambda}} \, \omega_{l,n=0} = \lam - i \frac{1}{2} +  \frac{7 - 61\eta }{216 L} 
- i \, \frac{137 + 1868\eta - 2005 \eta^2 }{7776L^2} \nn \\
\fl  &&   \quad \quad \quad
+ \frac{5230 + 440043\eta - 1274856\eta^2 + 750851\eta^3}{2519424 L^3}  \\
\fl &&
\quad \quad \quad + \, i \, \frac{590983 + 37791548 \eta - 243504102 \eta^2 + 340616636 \eta^3 - 135495065 \eta^4}{362797056 L^4}  + \mathcal{O}(L^{-5}) \nn
\end{eqnarray}
where $\eta = 9 \Lambda$. It is straightforward to extend to fields of higher spin, or to higher modes.
The results compare well with, e.g., the 6th-order WKB results of Zhidenko (\cite{Zhidenko}, Table 1). 

% Modifying the method to study black holes with `mixed' boundary conditions is 

\subsection{Nariai Space-time.}
Nariai space-time \cite{Nariai} is the natural metric on the Lorentzian version of $S^2 \times S^2$,
corresponding to $f(r)=g(r)= 1-r^2$ and $h(r)=1$. It provides a valuable check on our
ansatz since the QNM frequencies can be determined exactly; for a scalar field with curvature coupling
$V=V_\mathrm{eff}=\xi R = 4\xi$
\beq
\omega_{ln} =  \sqrt{L^2+ 4\xi -\textstyle{\frac12} } - i (n + \textstyle{\frac12}) \ .
\eeq
Our expansion precisely yields the large $L$ expansion of this exact result, so at least in this 
case our expansion is not merely asymptotic.

\subsection{Higher-Dimensional Black Hole.}

An uncharged spherically-symmetric higher-dimensional black hole in $D=d+2$ dimensions (i.e. one dimension of time and $d+1$ spatial dimensions) is known as a Schwarzschild-Tangherlini black hole \cite{Tangherlini}, and its metric is
\beq
ds^2 = -f(r) dt^2  +  f^{-1}(r) dr^2 + r^2 d \Omega_{(d)}^2
\eeq
where $f(r) = g(r) = 1 - (r_h / r)^{d-1}$, and $r_h$ is the outer horizon. There is a circular orbit at $r = r_c$ with critical impact parameter $b=b_c$ where
\beq
r_c = %\left( \frac{d+1}{2} \right)^{1/[d-1]} r_h \quad \quad \text{and} \quad \quad b_c =  r_c \, f^{-1/2} (r_c) .
\left( \frac{d+1}{2} \right)^{1/(d-1)} r_h \quad \quad \text{and} \quad \quad b_c =  \left( \frac{d+1}{d-1} \right)^{1/2} r_c \ .
\eeq 
The gravitational perturbations are divided into three classes, labelled `tensor', `vector' and `scalar' \cite{Kodama-Ishibashi, Konoplya-grav}. The tensor perturbations are the easiest to analyse; they satisfy the radial equation
\beq
\fl \quad \quad f(r) \frac{d}{d r} \left( f(r) \frac{d u}{d r} \right) + \left[  \omega^2 -  f(r) \left(  \frac{L^2 - (d-1)^2/4}{r^2} + \frac{f^\prime d}{2 r} + \frac{d(d-2)f(r)}{4} \right)  \right] u = 0
\eeq
where recall $L = l + (d-1)/2$. The lowest-order (in $L$) terms in the `tensor' QNM expansion are
\beq
\omega_{l, n=0} = \frac{L}{b_c} - i \frac{ \left(n + 1/2 \right) \sqrt{d - 1}}{b_c} + \mathcal{O} \left( L^{-1} \right)
\eeq
where we have used the result $r_c^2 k_c^\prime(r_c) = b_c \sqrt{d - 1}$. This result is, unsurprisingly, consistent with e.g.~Eq.~(51) in \cite{Lyapunov1} and Eq.~(12,13) in \cite{Konoplya}. It should be possible to apply the method to obtain higher-order terms, and it may be possible to analyse `vector' and `scalar' perturbations in the same way, although the potential for the `scalar' case is not positive definite, and it has more than one extrema \cite{Kodama-Ishibashi, Konoplya-grav, Zhidenko-2009}. %The QNMs of brane-localized Standard Model fields is also a subject under active investigation \cite{QNM-brane}.

\subsection{Canonical Acoustic Hole.\label{sec:acoustic}}
Black hole analogues -- systems with some of the key properties of black holes -- have received much attention in recent years \cite{Barcelo-Liberati-Visser, Unruh-Schutzhold}. One of the simplest models for an analogue in fluid mechanics is the `canonical acoustic hole', first proposed in \cite{Unruh}. An `horizon' forms where the bulk speed of the fluid (inwards) exceeds the speed of sound in the fluid. This raises the interesting possibility that QNM frequencies may one day be measured in the laboratory \cite{Berti-Cardoso-Lemos}. The effective geometry of the `canonical hole' is given by line element (\ref{le-ss}) with $f(r) = g(r) = 1 - r_h^4 / r^4$ and $h(r) = r^2$ (and $d=2$). The `fundamental' QNM frequencies are given by
\beq
\fl \quad \quad \quad b_c \omega_{l, n=0} = L - i - \frac{61}{216 L} - i \, \frac{17}{972 L^2} - \frac{532843}{2519424 L^3} + i \frac{4802843}{5668704 L^4} + \mathcal{O}(L^{-5})   \label{eq-acoustic}
\eeq
where $b_c = 3^{3/4} / 2^{1/2} r_h$ and $L = l+1/2$, a result which is consistent with Eq.~(53) in \cite{Berti-Cardoso-Lemos}. 
The accuracy of the QNMs obtained via (\ref{eq-acoustic}) is  comparable or superior to the best estimates in the literature. For example, at $l=2$, Eq.~(\ref{eq-acoustic}) gives $\omega r_h \approx 1.4726 - 0.6087 \pm [0.01 + 0.01i]$, which should be compared with $\omega r_h \approx 1.41 - 0.70 i$ found using a 6th-order WKB scheme \cite{Berti-Cardoso-Lemos}. A detailed study of the QNMs of acoustic holes is currently in progress \cite{Leandro}.

\section{Conclusion\label{sec-discussion}}

In this paper we have introduced a new method for obtaining QNM frequencies and wavefunctions through  expansion in inverse powers of the angular momentum $L$ (where $L = l + 1/2$ in four dimensions). We illustrated the method by applying to the Schwarzschild spacetime (Sec.~\ref{sec-schwarzschild}), to obtain an expansion of the QNM frequency (Eq.~\ref{omega-expansion} and Eq.~\ref{omco-0}--\ref{omco-5}) that may be taken to very high orders (Eq.~\ref{om-n0}, Tables \ref{table-n0} and \ref{table-n-higher}). We validated the frequency expansion against existing results (Table \ref{table-validation}), to show that it is accurate and rapidly convergent (Fig.~\ref{fig-ae-err}) in the regime $l \gtrsim 2$, $l \gtrsim n$ (but generally inaccurate/poorly convergent outside this regime, see Fig.~\ref{fig-ae-curves}). In Sec.~\ref{sec-regge} we showed that the expansion method may also be applied to find Regge poles (Eq.~\ref{lamco-0}--\ref{lamco-5}). In Sec.~\ref{sec-ss} we generalized the method to treat static spherically-symmetric spacetimes of arbitrary spatial dimension. The method was then applied to a selection of five such spacetimes of interest in Sec.~\ref{sec-examples}.

The expansion method complements the standard WKB approach \cite{WKB, Iyer-Will, Iyer-1987, Konoplya}. In addition, we believe the method holds certain key advantages. It provides a simple high-order expansion for QNM (and Regge pole) wavefunctions that is convergent everywhere outside the horizon (in the regime of validity $l \gtrsim 2, l \gtrsim n$); it yields the complex coefficient $\Aout$ (Eq.~\ref{Aout-n-gen} and \ref{Aout-n0}) which expresses the ratio of flux at infinity to flux at the horizon; and it provides additional insight into the link between orbiting null geodesics and QNMs. 

Let us briefly expand on this last point. The crucial step in the new method is a `geometric' ansatz for the wavefunction, given in Eq.~(\ref{ansatz}) and Eq.~(\ref{ansatz-general}). In Sec.~\ref{subsec:crit-orb} we showed how the ansatz is related to the family of `critical' null geodesics that approach from infinity and end in perpetual orbit on the photon sphere at $r=r_c$. In some sense, in this work we have conducted an expansion about these `critical' geodesics. In Sec.~\ref{subsec:wave-eq} we proved a key result: in the eikonal limit $l \rightarrow \infty$, the QNM frequencies depend only upon the properties of the `critical' null geodesics, confirming the result of \cite{Lyapunov1}. The real part of frequency is $\text{Re}(\omega_{ln})= L / b_c$, where $b_c $ is the critical impact parameter (and $b_c = 1/\Omega_c$ where $\Omega_c$ is the frequency of the null orbit at $r=r_c$), and the imaginary part is $\text{Im}(\omega_{l n}) = - (n+1/2) |\lambda|$, where $|\lambda|$ is the Lyapunov exponent given in Eq.~(\ref{lyap-def}). Our analysis is valid in general for any perturbation of a static spherically-symmetric spacetime that satisfies `radiative' conditions at both boundaries $\rstar \rightarrow \pm \infty$ (a condition which excludes, for example, Schwarzschild-AdS). %An advantage of our method is that higher-order terms in the expansion can be obtained easily, to move beyond the eikonal regime. 

In the highly-damped regime $n \gg l, n \gg 0$, it is well known that QNMs have the asymptotic form $\omega_{ln} \sim \Re - i n \Im$ (see e.g.~\cite{Andersson-Howls} for a discussion). Here $\Re$ and $\Im$ are real constants, where $\Re$ depends on the geometry and the spin of the field, and $\Im$ depends only on the geometry. For single-horizon asymptotically-flat black holes it has been shown that $\Im$ is set by the surface gravity of the horizon \cite{Padmanabhan-2004}; for multi-horizon holes the situation is more complicated \cite{Choudhury-Padmanabhan}. A possible challenge for the future would be to adapt the methods of this paper to investigate the highly-damped regime $n \gg l, n \gg 0$. 

It is hoped that the existence and properties (e.g.~mass, angular momentum) of astrophysical black holes may one day be inferred from observations of QNM ringing (in lightly-damped modes) at gravitational-wave detectors. We have seen that the relevant part of the QNM frequency spectrum is relatively easy to compute; however it is much more difficult to obtain an estimate of the degree of ringing excited by a given source. The theoretical framework has been in place for over twenty years \cite{Leaver-1986}, yet because of the technical complexity only a few attempts at practical calculations have been made \cite{Leaver-1986, Sun-Price, Nollert-Schmidt, Andersson-1997, Nollert-Price, Berti-Cardoso-2006}. This remains an interesting avenue for further study (and the Regge pole formalism, developed in \cite{Andersson-Thylwe-1994, CAM, Decanini-Folacci-Jensen, Decanini-Folacci-2009}, may provide an alternative route to the same objective), particularly in the present era of precision numerical relativity where linear perturbation theory estimates may be compared directly with non-linear exact results. A key ingredient in the recipe for computing the degree of QNM ringing is the so-called `QNM Excitation Factor': $\Bef = \Aout / ( 2 \omega \frac{\partial \Ain}{\partial \omega})$. In a forthcoming paper, we will show how the expansion method can be combined with WKB techniques to find $\partial \Ain / \partial \omega$ and to compute excitation factors in the large-$L$ limit. We will also show that the singular structure of the retarded Green function near the null cone may be deduce from the large-$L$ asymptotics of QNM wavefunctions and excitation factors \cite{CDOW1, Dolan-Ottewill-ef}.

Further extensions of this work are possible. Firstly, we hope it would be a simple task to use the expansion method to explore the examples in Sec.~\ref{sec-examples} (or others) in greater depth, or to study the Regge pole frequencies and wavefunctions for these spacetimes. Secondly, there remains the question of how the expansion method could be extended to treat the Schwarzschild-AdS case (of interest in the context of the AdS/CFT conjecture \cite{Nunez-Starinets}), where `reflecting' boundary conditions at $\rstar \to + \infty$ are imposed. Thirdly, how might the expansion method be applied to the Kerr (rotating black hole) spacetime? Here, an added complication is that the angular separation constant depends on the frequency of the perturbation. However, it is known that a one-parameter ($\mathcal{Q}$) family of unstable photon orbits at fixed (Boyer-Lindquist coordinate) $r=r_c(\mathcal{Q})$ exist \cite{Teo-2003}, so a geometrically-motivated line of enquiry may pay dividends. So far we have had some limited success in finding the large-$l$ asymptotics of the `polar' ($m=0$) mode of the scalar field. We hope to report on further progress in this direction in the near future.

\appendix

\ack 
%\begin{acknowledgements}
Thanks to Marc Casals and Barry Wardell for many insightful discussions which have influenced this work. S.D. thanks Leandro Oliveira and Luis Crispino for independently checking the canonical acoustic hole calculation (Eq.~\ref{eq-acoustic}). S.~D.~gratefully acknowledges the financial support of the Irish Research Council for Science, Engineering and Technology (IRCSET). 
%\end{acknowledgements}

\appendix

%\bibliography{year1}

\begin{thebibliography}{10}


\bibitem{Kokkotas-Schmidt}
K.~D.~Kokkotas and B.~G.~Schmidt,
% Quasi-normal modes of stars and black holes
Living Rev. Relativity {\bf 2} (1999) 2 % [http://relativity.livingreviews.org/Articles/lrr-1999-2/].

\bibitem{Nollert}
H-P. Nollert,
% Quasinormal modes: the characteristic `sound' of black holes and neutron stars
Class. Quantum Grav. {\bf 16} (1999) R159.

\bibitem{Ferrari-Gualtieri}
V.~Ferrari and L.~Gualtieri,
%Quasi-normal modes and gravitational wave astronomy.
Gen. Rel. Grav. {\bf 40} (2008) 1572 [arXiv:0709.0657].

\bibitem{Berti-Cardoso-Starinets}
E.~Berti, V.~Cardoso and A.~O.~Starinets,
%Quasinormal modes of black holes and black branes.
{\it Class.~Quantum Grav.} {\bf 26} (2009) 163001 [arXiv:0905.2975].

\bibitem{Vishveshwara}
C.~V.~Vishveshwara, Nature {\bf 227} (1970) 936.
% Scattering of gravitational radiation by a Schwarzschild black hole.
M. Davis, R. Ruffini, W. H. Press and R. H. Rice, Phys. Rev. Lett. {\bf 27} (1971) 1466.
%W. Press, Astrophys. J. {\bf 170} (1971) L105.

\bibitem{Chandrasekhar}
S. Chandrasekhar,
{\it The Mathematical Theory of Black Holes}
(Oxford University Press, New York, 1983).

\bibitem{Leaver-1986}
E.~Leaver, Phys.~Rev.~D {\bf 34} (1986) 384.

\bibitem{York}
J. W. York, Phys. Rev. D {\bf 28} (1983) 2929.

\bibitem{Hod}
% Bohr's Correspondence Principle and the Area Spectrum of Quantum Black Holes
S. Hod, Phys. Rev. Lett. {\bf 81} (1998) 4293 [gr-qc/9812002].

\bibitem{Dreyer}
O.~Dreyer, Phys.~Rev.~Lett. {\bf 90} (2003) 081301 [gr-qc/0211076].

\bibitem{Natario-Schiappa}
J.~Natario and R.~Schiappa,
Adv.~Theor.~Math.~Phys. {\bf 8} (2004) 1001 [hep-th/0411267].

\bibitem{Domagala-Lewandowski}
M.~Domagala and J.~Lewandowski,
% Black hole entropy from quantum geometry.
Class. Quantum Grav. {\bf 21} (2004) 5233 [gr-qc/0407051].

\bibitem{Nunez-Starinets}
A.~Nunez and A.~O.~Starinets,
Phys. Rev. D {\bf 67} (2003) 124013 [hep-th/0302026].

\bibitem{Evans-Threlfall}
N.~Evans and E.~Threlfall,
Phys. Rev. D {\bf 77} (2008) 126008 [arXiv:0802.0775].

\bibitem{Press}
W.~H.~Press,
Astrophys.~J.~{\bf 170} (1971) L105.

\bibitem{Goebel}
C.~J.~Goebel, 
Astrophys.~J.~{\bf 172} (1972) L95.

\bibitem{Ferrari-Mashhoon}
V.~Ferrari and B.~Mashhoon, 
Phys.~Rev.~D {\bf 30} (1984) 295.

\bibitem{Mashhoon}
B.~Mashhoon, Phys. Rev. D {\bf 31} (1985) 290.

\bibitem{Lyapunov1}
V.~Cardoso, A.~S.~Miranda, E.~Berti, H.~Witek and V.~T.~Zanchin,
% Geodesic stability, Lyapunov exponents and quasinormal modes.
Phys.~Rev.~D {\bf 79} (2009) 064016 [arXiv:0812.1806].

\bibitem{Lyapunov2}
L. Bombelli and E. Calzetta, Class. Quant. Grav. {\bf 9} (1992) 2573. 
 
\bibitem{Lyapunov3}
N. J. Cornish and J. J. Levin, Class. Quant. Grav. {\bf 20} (2003) 1649 [arXiv:gr-qc/0304056].

\bibitem{Decanini-Folacci-Jensen}
Y.~D\'ecanini, A.~Folacci and B.~Jensen,
% Complex angular momentum in black hole physics
Phys.~Rev.~D {\bf 67} (2003) 124017 [gr-qc/0212093].

\bibitem{Decanini-Folacci-2009}
Y.~D\'ecanini and A.~Folacci, (2009) 
% Regge poles of the Schwarzschild black hole: a WKB approach
[arXiv:0906.2601].

\bibitem{Berti}
E.~Berti,
% Black hole quasinormal modes: hints of quantum gravity?
%Proceedings from the ``Workshop on Dynamics and Thermodynamics of Black Holes and Naked SingularitiesÓ, Milan (2004) 
[gr-qc/0411025].

\bibitem{Dorband}
E.~N.~Dorband, E.~Berti, P.~Diener, E.~Schnetter and M.~Tiglio,
% A numerical study of the quasinormal mode excitation of Kerr black holes
Phys.~Rev.~D {\bf 74} (2006) 084028 [gr-qc/0608091].

\bibitem{Chandrasekhar-Detweiler}
S. Chandrasekhar and S. Detweiler, Proc. Roy. Soc. London A {\bf 344} (1975) 441.

\bibitem{Blome-Mashhoon}
H.-J.~Blome and B.~Mashhoon, Phys.~Lett.~{\bf 100A} (1984) 231. 

\bibitem{WKB}
B.~F.~Schutz and C.~M.~Will, Astrophys.~J.~Lett. {\bf 291} (1985) L33.

\bibitem{Iyer-Will}
S.~Iyer and C.~M.~Will, Phys.~Rev.~D {\bf 35} (1987) 3621.

\bibitem{Iyer-1987}
S.~Iyer, Phys.~Rev.~D {\bf 35} (1987) 3632.

\bibitem{Konoplya}
R.~A.~Konoplya, Phys.~Rev.~D {\bf 68} (2003) 024018 [gr-qc/0303052].
% Quasinormal behavior of the D-dimensional Schwarzschild black hole and the higher order WKB approach

\bibitem{Phase-integral1}
N. Andersson, Proc.~R.~Soc.~A {\bf 439} (1992) 47.

\bibitem{Phase-integral2}
N.~Andersson and S. Linnaeus,
Phys. Rev. D {\bf 46} (1992) 4179.

\bibitem{Leaver-1985}
E.~W. Leaver, Proc.~R.~Soc.~A {\bf 402} (1985) 285.

\bibitem{Nollert-1993}
H.-P. Nollert, Phys. Rev. D {\bf 47} (1993) 5253.

\bibitem{BenderOrzag}
C.~M.~Bender and S.~A.~Orzag,
{\it Advanced Mathematical Methods for Scientists and Engineers} (Springer, New York, 1999).

\bibitem{Dolan-Ottewill-ef}
S.~R.~Dolan and A.~C.~Ottewill,
{\it in preparation}.

%%%%%%%%%%%%%%%%%%%%%%%%
% Compex Angular Momentum / Regge poles
%%%%%%%%%%%%%%%%%%%%%%%%

\bibitem{CAM}
R.~G.~Newton, \emph{Scattering Theory of Waves and Particles} (Springer-Verlag, New York, 2nd Ed.~1982).

\bibitem{Collins}
P. D. B. Collins, \emph{An Introduction to Regge Theory and High-Energy Physics}, Cambridge University Press (Cambridge, UK, 1977).

\bibitem{Andersson-Thylwe-1994}
N.~Andersson and K.-E.~Thylwe,
% Complex angular momentum approach to black-hole scattering
Class. Quantum Grav. {\bf 11} (1994) 2991.

\bibitem{Andersson-1994}
N.~Andersson,
%Complex angular momenta and the black-hole glory
Class. Quantum Grav. {\bf 11} (1994) 3003.

\bibitem{Andersson-Jensen}
N.~Andersson and B.~Jensen,
% Scattering by black holes
[gr-qc/0011025].

%F.~W.~J.~Olver,
%{\it Asymptotics and Special Functions} (Academic Press, New York, 1974).

%\bibitem{Kay-Radzikowski}
%Kay, B.S. and Radzikowski, M.J. and Wald, R.M.,
%\textit{Quantum Field Theory on Spacetimes with a Compactly Generated Cauchy Horizon},
%Commun. Math. Phys., \textbf{183}, 533 (1997).


%\bibitem{CDOW1}
%M.~Casals, S.~R.~Dolan, A.~C.~Ottewill and B.~Wardell.
% 
%Phys. Rev. D, {\bf 79}, 124043 (2009) [arXiv:0903.0395].

%\bibitem{Aki-Richards}
%K. Aki and P. G. Richards, Quantitative Seismology (University Science Books, 2002).

%\bibitem{Andersson-1997}
%N.~Andersson. 
%Phys. Rev. D {\bf 55} 468 (1997).

%\bibitem{Perlick}
%V.~Perlick.
%%"Gravitational Lensing from a Spacetime Perspective",
%Living Rev. Relativity {\bf 7}, (2004), 9. http://www.livingreviews.org/lrr-2004-9 

%\bibitem{Ori1}
%A.~Ori (2008). 
%The four-fold structure of the singular part of the Green
%  function beyond the caustics was found by Ori some time ago. He also
%  conducted measurements in an analog acoustic system, which seem to verify his
%  theoretical prediction. Private communication (2008) and report (2009)
%  available at {http://physics.technion.ac.il/\~{}amos/acoustic.pdf}.


\bibitem{Berti-Cardoso-2006}
E.~Berti and V.~Cardoso,
% Quasinormal ringing of Kerr black holes: The excitation factors.
Phys.~Rev.~D {\bf 74} (2006) 104020 [gr-qc/0605118].


%%%%%%%%%%%%%%%%%%%
% Reissner-N\"ordstrom black holes
%%%%%%%%%%%%%%%%%%%
\bibitem{Leaver-1990}
E.~W.~Leaver, 
Phys.~Rev.~D {\bf 41} (1990) 2986.

\bibitem{Andersson-Araujo-Schutz}
N.~Andersson, M.~E.~Ara\'ujo and B.~F.~Schutz,
Phys.~Rev.~D {\bf 49} (1994) 2703.

\bibitem{Onozawa}
% Quasinormal modes of maximally charged black holes
H. Onozawa, T. Mishima, T. Okamura and H. Ishihara, 
Phys.~Rev.~D {\bf 53} (1996) 7033 [gr-qc/9603021].

%%%%%%%%%%%%%%%%%%%
% Schwarzschild-de Sitter black holes
%%%%%%%%%%%%%%%%%%%

\bibitem{Zhidenko}
A.~Zhidenko, 
% Quasi-normal modes of SchwarzschildÐde Sitter black holes
Class.~Quantum Grav. {\bf 21} (2004) 273 [gr-qc/0307012].

%%%%%%%%%%%%%%%%%%%
% Higher-dimensional black holes
%%%%%%%%%%%%%%%%%%%
\bibitem{Tangherlini}
F.~R.~Tangherlini, Nuovo Cim. {\bf 27} (1963) 636.

\bibitem{Kodama-Ishibashi}
H.~Kodama and A.~Ishibashi, Prog.~Theor.~Phys. {\bf 110} (2003) 701 [hep-th/0305147]. \\
A.~Ishibashi and H.~Kodama, Prog.~Theor.~Phys. {\bf 110} (2003) 901 [hep-th/0305185]. \\
H.~Kodama and A.~Ishibashi, Prog.~Theor.~Phys. {\bf 111} (2004) 29 [hep-th/0308128].

\bibitem{Konoplya-grav}
% Gravitational quasinormal radiation of higher-dimensional black holes
R.~A.~Konoplya, Phys.~Rev.~D {\bf 68} (2003) 124017 [hep-th/0309030].

\bibitem{Zhidenko-2009}
A.~Zhidenko, 
{\it Linear perturbations of black holes: stability, quasi-normal modes and tails}. Ph.D thesis, Universidade de S\~ao Paulo (2009) [arXiv:0903.3555].

%%%%%%%%%%%%%%%%%%%
% The Nariai spacetime
%%%%%%%%%%%%%%%%%%%
\bibitem{Nariai}
H.~Nariai, Sci. Rep. Tohoku Univ. {\bf 34} (1950) 160.
H.~Nariai, Sci. Rep. Tohoku Univ. {\bf 35} (1951) 62.

%%%%%%%%%%%%%%%%%%%
% Canonical acoustic holes
%%%%%%%%%%%%%%%%%%%

\bibitem{Barcelo-Liberati-Visser}
C.~Barcel\'o, S.~Liberati, and M.~Visser, {\it Living Rev.~Relativity} {\bf 8} (2005) 12 [http://relativity.livingreviews.org/Articles/lrr-2005-12/] [gr-qc/0505065].

\bibitem{Unruh-Schutzhold}
W.~G.~Unruh and R.~Schutzhold, {\it Quantum Analogues: From Phase Transitions to Black Holes and Cosmology}, Lect.~Notes Phys.~Vol.~{\bf 718} (Springer, Berlin, 2007).

\bibitem{Unruh}
W. G. Unruh, Phys.~Rev.~Lett.~{\bf 46} (1981) 1351.

\bibitem{Berti-Cardoso-Lemos}
E.~Berti, V.~Cardoso and J.~P.~S.~Lemos,
%Quasinormal modes and classical wave propagation in analogue black holes
Phys.~Rev.~D {\bf 70} (2004) 124006 [gr-qc/0408099].

\bibitem{Leandro}
L.~A.~Oliveira and L.~C.~B.~Crispino, {\it in preparation}.

\bibitem{CDOW1}
M.~Casals, S.~R.~Dolan, A.~C.~Ottewill and B.~Wardell, 
Phys. Rev. D {\bf 79} (2009) 124043 [arXiv:0903.0395].

%%%%%%%%%%%%%%%%%%%
% QNM Excitations
%%%%%%%%%%%%%%%%%%%

\bibitem{Sun-Price}
Y.~Sun and R.~H.~Price, Phys.~Rev.~D {\bf 38} (1988) 1040.

\bibitem{Nollert-Schmidt}
H.-P. Nollert and B.G. Schmidt, Phys.~Rev.~D {\bf 45} (1992) 2617.

\bibitem{Andersson-1997}
N.~Andersson, Phys.~Rev.~D {\bf 55} (1997) 468 [gr-qc/9607064].

\bibitem{Nollert-Price}
H.-P.~Nollert and R.~H.~Price, J.~Math.~Phys. {\bf 40} (1999) 980 [gr-qc/9810074].

%%%%%%%%%%%%%%%%%%%
% Discussion of large-n limit
%%%%%%%%%%%%%%%%%%%
\bibitem{Andersson-Howls}
N.~Andersson and C.~J.~Howls,
% The asymptotic quasinormal mode spectrum of non-rotating black holes
Class.~Quantum Grav. {\bf 21} (2004) 1623 [gr-qc/0307020]. 

\bibitem{Padmanabhan-2004}
T.~Padmanabhan,
% Quasi normal modes: A simple derivation of the level spacing of the frequencies,
Class.~Quantum Grav. {\bf 21} (2004) L1 [gr-qc/0310027].

\bibitem{Choudhury-Padmanabhan}
T.~Roy Choudhury and T.~Padmanabhan,
% Quasi normal modes in Schwarzschild-DeSitter spacetime: A simple derivation of the level spacing of the frequencies.
Phys.~Rev.~D {\bf 69} (2004) 064033 [gr-qc/0311064].

\bibitem{Teo-2003}
E.~Teo,
% Spherical photon orbits around a Kerr black hole
Gen.~Rel.~Grav.~{\bf 35} (2003) 1909.

\end{thebibliography}

\section*{References}
\bibliographystyle{unsrt}

\end{document}